\pdfoutput=1

\documentclass[english,aps,prd,superscriptaddress,preprintnumbers,floatfix,nofootinbib,12pt]{revtex4-1}
\usepackage{amsmath,amssymb,amsfonts}
\usepackage{amsmath}
\usepackage{graphicx}

\usepackage{hyperref}
\hypersetup{colorlinks,linkcolor={blue},citecolor={blue},urlcolor={blue}} 
\usepackage{braket}
\usepackage[english]{babel}
\usepackage{bm}
\usepackage[utf8]{inputenc}

\usepackage{listings}

\usepackage[T1]{fontenc}
\usepackage{lmodern}
\usepackage{caption}
\usepackage{floatrow}

\usepackage{subcaption}
\usepackage{mathrsfs}
\usepackage{enumerate}

\usepackage[normalem]{ulem}
\usepackage{color}
\definecolor{nicered}{rgb}{0.8,0.15,0.15}

\def\beq{\begin{equation}}
\def\eeq{\end{equation}}
\def\be{\begin{equation}}
\def\ee{\end{equation}}
\def\bea{\begin{eqnarray}}
\def\eea{\end{eqnarray}}

\begin{document}

\title{European Spallation Source as a searching tool for an ultralight scalar field}   
\author{Rub\'en Cordero}\email{rcorderoe@ipn.mx} \affiliation{Departamento de F\'{\i}sica, Escuela Superior de
  F\'{\i}sica y Matem\'aticas del Instituto Polit\'ecnico Nacional, Unidad Adolfo L\'opez Mateos, Edificio 9, 07738 Ciudad de M\'exico, Mexico}
\author{Luis A. Delgadillo} \email{ldelgadillof2100@alumno.ipn.mx}
\affiliation{Departamento de F\'{\i}sica, Escuela Superior de
  F\'{\i}sica y Matem\'aticas del Instituto Polit\'ecnico Nacional, Unidad Adolfo L\'opez Mateos, Edificio 9, 07738 Ciudad de M\'exico, Mexico}
\author{O. G. Miranda} \email{omar.miranda@cinvestav.mx}
\affiliation{Departamento de F\'{\i}sica, Centro de Investigaci{\'o}n y de Estudios Avanzados del IPN
  Apdo. Postal
  14-740 07000 Ciudad de M\'exico, Mexico}

\begin{abstract}
\noindent
Dark matter (DM) nature is one of the major issues in
physics. The search for a DM candidate has motivated the known proposal of an ultralight scalar
field (ULSF). We explore the possibility
to search for this ULSF at the upcoming European Spallation
Source neutrino Super-Beam experiment. We have
considered the recent study case in which there could be an
interaction between the ULSF and active neutrinos. We have found that
in this future experimental setup, the sensitivity is
competitive with other neutrino physics experiments. We show the
expected future sensitivity for the main parameter modeling the
interaction between ULSF and neutrinos.
\end{abstract}

\maketitle





\section{Introduction}
\label{sec:intro}

The existence of dark matter (DM) is one of the most intriguing aspects of
modern cosmology. Nearly one fourth of the matter-energy content of the
Universe is in the form of dark matter and elucidating its nature is a
very fundamental issue. There are plenty of proposals to model dark
matter, from objects like primordial black holes with masses of the
order of $2\times 10^{21}$ Kg ($10^{57}$ eV/$c^2$), weakly
interacting massive particles (WIMPs)  with masses around $100$ TeV/$c^2-$2
GeV/$c^2$, axions with masses of $10^{-5}$ eV/$c^2$, or ultralight
scalar dark matter particles with masses of $10^{-22}$ eV/$c^2$.

One of the proposals for a dark matter candidate postulates that an
ultralight relativistic scalar field $\phi$ can model dark matter
with a suitable scalar potential $V(\phi)$
\cite{Sahni:1999qe,Hu:2000ke,Matos:1999et,Matos:2000ng,Arbey:2001jj,Arbey:2001qi}
(see, for example, the following reviews
\cite{Magana:2012ph,Suarez:2013iw,Marsh:2015xka,Hui:2016ltb,Lee:2017qve,Urena-Lopez:2019kud}). The
general property of this scalar field potential is the existence of a
parabolic minimum around which it is possible to define a mass scale
for the related boson particle. In this model, the galactic halos can
be formed by condensing the scalar field at the Universe's beginning
\cite{Magana:2012ph}. The successful physical implications of
$\Lambda$CDM at cosmological scales are replicated by the scalar field
dark matter, e.g., the evolution of the cosmological densities
\cite{Matos:2008ag}, the acoustic peaks of the cosmic microwave
background radiation (CMBR) \cite{Rodriguez-Montoya:2010ebl}, the
rotations curves in big- and low-surface brightness (LSB) galaxies
\cite{Lesgourgues:2002hk,Arbey:2003sj,Harko:2011jy,Robles:2012uy}, and
the observed properties of dwarf galaxies \cite{Lee:2008jp}.

Besides, the oscillations of the scalar field could have exciting and
important physical consequences. Since neutrinos have small masses,
the interaction with the oscillations of the scalar field could have a
relevant influence on neutrino masses and mixing. Considering that
the mechanism of neutrino masses is still unknown, the interaction
with the oscillating scalar field could provide a link to physics
beyond the Standard Model (SM). The former reasons constitute a strong
motivation to study the effects of a light scalar field in neutrino
oscillations.

More recently, searches for an ultralight scalar field at neutrino
oscillation experiments are becoming possible (see, e.g.,
Refs.~\cite{Berlin:2016woy,Brdar:2017kbt,Krnjaic:2017zlz,Liao:2018byh,
  Kim,Dev:2020kgz,Losada:2021bxx}) due to advancements in the energy
resolution of present and upcoming neutrino oscillation
experiments. Besides, feasible cosmological models of active and/or
sterile neutrinos interacting with an ultralight scalar field have
been explored
\cite{Zhao:2017wmo,Farzan,Cline:2019seo,Dev:2022bae,Huang:2022wmz}. In
Ref.~\cite{Losada:2022uvr}, the implications for the reactor neutrino
experiments Kamioka liquid-scintillator antineutrino detector
(KamLAND) and Jiangmen Underground Neutrino Observatory (JUNO) due to
interactions between active neutrinos and a scalar field with mass
$m_{\phi} \gtrsim 10^{-12}$ eV/$c^2$ are
considered. Despite these interesting studies, it is
important to notice that such an interaction between neutrinos
and the ultralight scalar field (ULSF) may also lead to tension with current experimental
observables, as we will discuss below. Moreover, recent studies corner the ultralight dark matter (ULDM) proposal
to be only a subdominant component of the DM content \cite{Kobayashi:2017jcf,Bar:2021kti}. Still, the interaction of a neutrino with a ULSF can be of interest independently of its connection to the DM problem.

In this paper, we will explore the phenomenological consequences of
having an ultralight scalar field mixed with the neutrino mass
eigenstates and the capabilities of an European Spallation Source Neutrino Super Beam (ESS$\nu$SB)-like
experiment~\cite{ESSnuSB:2013dql, ESSnuSB:2021azq, Alekou:2022emd} to constraint these
interactions. The original proposal is to study a neutrino Super Beam,
which employs the European Spallation Source (ESS) facility as a
neutrino source with a water Cherenkov detector \cite{MEMPHYS:2012bzz}
located in a deep mine, for the discovery of the Dirac $CP$-violating
phase $\delta$. We consider two baselines. One at 360 km,
corresponding to the distance from the source to the Zinkgruvan
mine. While the second would be at 540 km, with a detector placed at
the Garpenberg mine; both mines are located in Sweden. Moreover,
scenarios that investigate the capabilities of the ESS$\nu$SB
experiment to probe physics beyond the Standard Model and neutrino
oscillations have been discussed
\cite{KumarAgarwalla:2019blx,Ghosh:2019zvl,Choubey:2020dhw,
  Majhi:2021api,Blennow:2020snb, Chatterjee:2021xyu,Ahn:2022ufs}.

The structure of the paper is as follows. In Sec.~\ref{frame}, we
present one possible form of interaction between neutrinos and the
ultralight scalar field. This interaction modifies the leptonic
mixing angle $\theta$, and adds a smearing on the neutrino mass
squared difference $\Delta m^2$. Section~\ref{simulation} explains
the characteristics and assumptions made in the general long
baseline experiment simulator (GLoBES)
software~\cite{Huber:2004ka,Huber:2007ji} to simulate the ESS$\nu$SB
experiment. Sensitivities to the ULSF via modulations from average
distorted neutrino oscillations are developed in
Sec.~\ref{mod}. Finally, we give our conclusions.

\section{Framework}
\label{frame}

The existence of dark matter has stimulated extensive and intensive
activity to explain its characteristics. There are a plethora of
possible candidates for dark matter. In this section, we will mention
some of the more studied proposals. For example, primordial black
holes could have been formed soon after the big bang from the gravitational
collapse of higher-density mass regions. Some constraints restrict the
masses of primordial black holes to several windows between $
10^{13}-10^{14}$ kg, $10^{17}-10^{21}$ kg, and $10-10^3$ solar masses
($2\times 10^{30}$ kg) \cite{Carr:2020xqk}. However, from latest
results of LIGO and VIRGO, it seems that primordial black holes only
provide some part of the needed amount of dark matter
\cite{Carr:2020mqm}.

The main characteristics that must fulfill a possible particle
candidate for dark matter are that it has to be stable over billions
of years, nonrelativistic, massive, and weakly interacting. The
Standard Model of particle physics does not have a particle with these
properties.

One of the most studied extensions of the Standard Model is its
minimal supersymmetric extension (MSSM) \cite{Arbey:2021gdg}; several
candidates for WIMPs can emerge in this model. The possible candidates
for dark matter are neutralinos, gravitinos, and sneutrinos. The
neutralino is the most studied particle candidate for dark matter; it
has been extensible searched at the LHC. In the LEP and Tevatron
experiments, a lower-mass bound around 46 GeV/$c^2$ has been set
\cite{Arbey:2021gdg}. The gravitino couples very weakly to other
particles; therefore, it is challenging to impose any constraint on
it~\cite{Arbey:2021gdg}. The lightest sneutrino is strongly
interacting, which is unsuitable for a dark matter particle.

Another popular dark matter candidate is the axion, a light neutral
particle that can be produced in the early Universe by a spontaneous
symmetry breaking of $U_a(1)$ Peccei-Quin symmetry
\cite{Arbey:2021gdg,DiLuzio:2020wdo}. Experimental attempts have been
developed to detect the axions using the prediction that axions and
photons could be transformed into each other in an intense magnetic
field \cite{Arun:2017uaw}.

Besides the former candidates for dark matter, there are other massive
particles like sterile neutrinos that only interact gravitationally,
with masses around
$\mathcal{O}$(keV/$c^2$)~\cite{Drewes:2013gca,Ng:2019gch}.

Furthermore, other exotic dark matter candidates exist like WIMPzillas, strongly interacting massive particles (SIMPs), Q-nuggets, Q-balls, gluinos, Fermi balls, EW balls, GUT balls, etc. The masses of these objects range from 100 GeV/$c^2$ to a TeV/$c^2$ \cite{Arun:2017uaw,Oks:2021hef}.

Another approach consists of avoiding the existence of massive dark
matter particles or objects and instead considering modifications of
gravitational interactions, for example, modified newtonian
dynamics (MOND)~\cite{Milgrom:1983ca,Baker:2019gxo} and extra
dimensions~\cite{Randall:1999ee,Randall:1999vf}.

Regarding the scalar-field dark matter, there are some problems that
this proposal can solve. For example, on the cosmological side there
are problems with certain predictions of $\Lambda$CDM at the galactic
scale. Some examples are the excess of substructures produced in
$N$-body numerical simulations, which are one order of magnitude larger
than the observed ones, and the cusp profile of central density in
galactic halos
\cite{Clowe:2006eq,Klypin:1999uc,Moore:1999nt}. Additionally, problems
arise in numerical simulations of structure formation, which do not
produce pure disk galaxies, among other problems. The former problems
could be avoided if the structure grew faster than in $\Lambda$CDM
\cite{Peebles:2010di}.

Some of the former problems are resolved in the ultralight
relativistic scalar field framework. In this model, the galactic halos
are formed by a Bose-Einstein condensation of a scalar boson with a
mass around $m_{\phi} \sim 10^{-22}$ eV/$c^2$. 

 The Compton length associated
with this boson is of the order of kpc, which is the same order as the
size of dark halos in typical galaxies. 
It is proposed that the dark
halos are very big drops of scalar field. Then, when the Universe
reaches the critical temperature of condensation, all galactic halos
form at the same time producing well-formed halo galaxies at high $z$,
which is a different prediction from $\Lambda$CDM \cite{Lee:2008jp}.

The scalar field dark matter can resolve the problem of cusp profile
of density in galactic halos since this is avoided due to the wave
properties of the ultralight mass of the scalar particles
\cite{Hu:2000ke,Lundgren:2010sp}. Furthermore, the excess of
substructures is prevented by considering that the scalar field has a
natural cutoff \cite{Hu:2000ke,Matos:2000ss,Suarez:2011yf}.

Although we consider here a scalar boson mass, $m_{\phi} \sim 10^{-22}$ eV/$c^2$, it is important to mention that this value can change from different physical constraints and different astrophysical models. Except for some few models \cite{Arvanitaki:2009fg, Cicoli:2021gss}, all the justification for this mass region comes from consistency with astrophysical observations. For example, from the anisotropies of the CMB, the mass of the scalar field could be in the range of  $m_{\phi} = 10^{-24}- 10^{-22}$ eV/$c^2$ \cite{Urena-Lopez:2019kud}. From galaxy rotation curves, the ULDM mass lies in the region $m_{\phi} = 0.5 \times 10^{-23}- 10^{-21}$ eV/$c^2$~\cite{Bernal:2017oih,Urena-Lopez:2017tob}.

In addition, it is expected that the ULDM cannot be the total fraction of the dark matter if the mass is light enough, although the exact value might be model dependent. For instance, the combined analysis of CBM and large-scale structure (LSS) data sets~\cite{Hlozek:2014lca} allow ULDM masses $m_{\phi} \gtrsim 10^{-24}$ eV/$c^2$ if $\rho_{\phi}\sim \rho_{\text{DM}}$ and as low as $m_{\phi} \sim 10^{-27}$ eV/$c^2$ if $\rho_{\phi} \sim 0.05 \rho_{\text{DM}}$. Lyman-$\alpha$ forest excludes ULDM masses lighter than $m_{\phi} \lesssim 10^{-21}$ eV/$c^2$ if $\rho_{\phi}\sim \rho_{\text{DM}}$ while a ULDM mass $m_{\phi} \sim 10^{-22}$ eV/$c^2$ can be accommodated if $\rho_{\phi} \lesssim 0.2 \rho_{\text{DM}}$ \cite{Kobayashi:2017jcf}. 
On the other hand, from the soliton-halo relation, a ULDM with mass below $m_\phi \sim 10^{-21}$ eV/$c^2$ is disfavored \cite{Bar:2018acw}. Moreover, Ref.~\cite{Bar:2021kti} suggests that the ULSF is not the total component of cosmological DM, if the mass range is $10^{-24}~ \lesssim m_{\phi}~(\text{eV}/c^2)~\lesssim 10^{-20}$, while admitting it for $\rho_{\phi} \lesssim 0.3 \rho_{\text{DM}}$.
Finally, constraints from structure formation \cite{Arvanitaki:2014faa} exclude ULDM masses lighter than $m_{\phi} \sim 10^{-20}$ eV/$c^2$ if $\rho_{\phi}\sim \rho_{\text{DM}}$ while the bound disappears when $\rho_{\phi} \lesssim 0.1 \rho_{\text{DM}}$.

Let us review some of the main characteristics of the scalar field framework, which are relevant to our work. From the
Lagrangian density for the scalar field
\begin{equation}
    \mathcal{L} = \frac{1}{2}\nabla_\mu \phi \nabla^\mu \phi + V(\phi),
\end{equation}
the conservation of the energy-momentum tensor $\nabla_\mu T^{\mu \nu}=0$ in the cosmological background of the Friedmann-Lemaitre-Robertson-Walker metric gives
\begin{equation}
    \ddot{\phi}+3H\dot{\phi}+V^{\prime}(\phi)=0,
    \label{fievolution}
\end{equation}
where $H = \dot{a}/a$ is the Hubble parameter and $a$ is the scale
factor of the Universe. In the following, we will use natural units
where $c=\hbar=1$. The scalar energy density is \cite{Magana:2012ph}
\begin{equation*}
    T_{0}^0= \frac{1}{2} \dot{\phi}^2+V(\phi)=\rho_{\phi}.
\end{equation*}
The evolution of the scalar field can be obtained numerically near the
minimum of the potential where $V^{\prime}(\phi) = m^2_\phi \phi$. The
contribution from the other components of matter-energy density
present in the Universe is included in the Friedmann equation
\begin{equation}
    H^2 = \frac{8\pi G}{3}(\rho_r+ \rho_m+ \rho_\phi + \rho_\Lambda),
\end{equation}
and $\rho_r$, $\rho_m$, and $\rho_\Lambda$ are the energy densities
associated with radiation, baryonic nonrelativistic matter, and dark
energy, respectively. However, there are analytical approximations for
the scalar fields in recent times when $H_0 \sim 10^{-33}$ eV and $H_0
\ll m_\phi$. It has been proposed \cite{Magana:2012ph} the following
anzats for the scalar field
\begin{equation}
\label{rhophi}
    \phi= 2\sqrt{\hat{\rho}}\cos(S-m_{\phi} t) \,\, .
\end{equation}
In late times, the scalar field behaves as nonrelativistic, and the
relation $\dot{S}/m_{\phi} \sim 0$ is fulfilled when the temperature
at which the scalar field begins to oscillate is $T_{\text{osc}} \sim$
keV corresponding to a redshift $z_{\text{osc}} \sim
10^6$~\cite{Hui:2016ltb} if $m_{\phi} \sim 10^{-22}$ eV.

From the evolution equation for the scalar field
Eq.~({\ref{fievolution}}), $\hat{\rho} = \hat{\rho}_0
a^{-3}$ is obtained, which is proportional to the energy density of
non-relativistic matter. From the expression of the scalar-field
energy density, it is possible to obtain $\rho_\phi = 2m^2_\phi
\hat{\rho}$. The ultralight scalar field can be described by a classical field minimally coupled to
gravity~\cite{Urena-Lopez:2007cpl}. For instance, by setting the phase
$S=\pi/2$ and writing the scalar field in terms of the energy density,
we can have a ULSF that oscillates with time as
\begin{equation}
\label{phi}
    \phi\simeq \phi_{0} \sin (m_{\phi}t), 
\end{equation}
with $m_{\phi}\sim 10^{-22}$ eV and
\begin{equation}
\label{phinot}
    \phi_{0} \simeq \frac{\sqrt{2\rho_\phi} }{m_\phi},
\end{equation}
where $\rho_\phi$ is the field density at the surface of the Earth,
which we will assume to be $0.3$~GeV/cm$^3$, $\rho_{\phi} \leq \rho_{\text{DM}, \odot} = 0.3$~GeV/cm$^3$. From recent analyses, the estimation of the local DM density coincide within a range of $\rho_{\text{DM}, \odot} \simeq 0.3-0.6~ \text{GeV/cm}^3$ \cite{deSalas:2019pee,deSalas:2020hbh,Sivertsson:2022riu}. 

If we consider that this scalar field can interact with the neutrino,
a modification in the leptonic mixing angle $\theta$ or additional
smearing on the neutrino mass-squared difference $\Delta m^2$ will
arise. Recently, the possible interaction between the ULSF and neutrinos and its implications have been
considered~\cite{Berlin:2016woy,Farzan,Dev:2020kgz,Losada:2021bxx}. In
this case, the ULSF can produce an effect on neutrino
oscillations. Besides the SM Lagrangian, we would have an additional
contribution due to the hypothetical ULSF interaction with the
neutrino~\cite{Berlin:2016woy,Losada:2021bxx}
\begin{equation}
\label{inter}
    \mathcal{L}_{\lambda,y}  \supset  \frac{\lambda^{\alpha \beta} }{\Lambda}  (L_{\alpha})^{\text{T}} L_{\beta} H H+ \frac{y^{\alpha \beta} }{\Lambda^{2}} \phi (L_{\alpha})^{\text{T}} L_{\beta} H  H,
\end{equation}
where $\lambda^{\alpha \beta}$ and $y^{\alpha \beta}$ are 3$\times$3
dimensionless symmetric matrices, and $\Lambda$ is the scale of new
physics.
\begin{equation}
\label{inter2}
    \mathcal{L}_{m_{\nu}}  \supset  \frac{\lambda^{\alpha \beta} v^2 }{2 \Lambda}  (\nu_{\alpha})^{\text{T}} \nu_{\beta}  + \frac{y^{\alpha \beta} v^2  }{2  \Lambda^{2}} \phi (\nu_{\alpha})^{\text{T}} \nu_{\beta}. 
\end{equation}
After symmetry breaking and replacement of the Higgs field $H$ by its
vacuum expectation value $\braket{H} = 1 / \sqrt{2} ~(0,v)^T$, the neutrino mass matrix
acquires corrections from the ULSF field $\phi$:
\begin{equation}
\label{int}
\Tilde{m}= m_{\nu}+\hat{y} \phi;~~ m_{\nu} = \frac{\lambda v^2}{\Lambda}~~\text{and}~~
    \hat{y}=  \frac{y   v^2  }{\Lambda^{2}}.
\end{equation}

For instance, the interaction term in Eqs.~(\ref{inter}) and (\ref{inter2}) can arise within the framework of a type I seesaw \cite{Krnjaic:2017zlz}.

If we consider the off-diagonal $\hat{y}$ matrix
elements, we will obtain a correction to the leptonic mixing angle. In
a $2\times2$ neutrino picture, the mixing matrix for this case will
have the form
\begin{equation}
\Tilde{m}= m_{\nu}+\hat{y} \phi = \left( \begin{array}{cc}
m_{1} & \hat{y}_{12} \phi  \\
 \hat{y}_{12} \phi & m_{2} \\
 \end{array} \right).
\end{equation}

\vspace{2.0mm}
To diagonalize this matrix, we can apply a rotation,
$R(\psi)$, such that $ \tan(2 \psi) = - 2 \hat{y}_{12}\phi / \Delta
m$. For small angles, $\psi$, $\tan(2 \psi) \approx 2 \psi $ and $
\psi \approx - \hat{y}_{12}\phi / \Delta m$. Once the mass matrix is
diagonal, we can rotate to the flavor basis through a new
transformation $R(\theta)$. Since $R(\theta)R(\psi) = R( \theta +
\psi)$, then, the mixing angle receives contributions from the
$\hat{y}_{12}$ term, such that $\Tilde{\theta} \rightarrow \theta +
\hat{y}_{12}\phi / \Delta m,~\Tilde{\theta} \rightarrow \theta +
\eta_{\theta} \sin m_{\phi} t$. However, our sensitivity, in this
case, will be limited. Therefore, we will concentrate on the case of
diagonal couplings.

In the case of two neutrino mixing, if we consider only diagonal
couplings~($\alpha=\beta$), the modified $\hat{y}$ matrix up to
leading order is
\begin{equation}
\Tilde{m}^2=( m_{\nu}+\hat{y} \phi)^2 \simeq \left( \begin{array}{cc}
m_{1}^2+2 m_{1}  \hat{y}_{11} \phi & 0  \\
 0 & m_{2}^2 + 2 m_{2} \hat{y}_{22} \phi   \\  
\end{array} \right) +  \mathcal{O}(\hat{y}^2\phi^2). 
\end{equation}
Therefore, for the mass squared difference, we will have 
\begin{equation}
\Delta \Tilde{{m}}^2_{21}  \simeq  m_{2}^2-m_{1}^2 + 2 (m_{2} \hat{y}_{22}- m_{1}  \hat{y}_{11} )\phi.
\end{equation}
or
\begin{equation}
\Delta \Tilde{m}^2_{21} = \Delta m_{21}^2 \Big[1+\frac{2 (m_{2} \hat{y}_{22}- m_{1}  \hat{y}_{11} )\phi}{\Delta m_{21}^2} \Big] + \mathcal{O}(\hat{y}^2\phi^2)  = \Delta m_{21}^2 \Big[1+ 2\eta_{\Delta} \sin(m_{\phi}t) \Big] + \mathcal{O}(\hat{y}^2\phi^2). 
\end{equation} 
Furthermore, the ULSF parameter $\eta_{\Delta}$ is given by
\begin{equation}
\label{eta}
   \eta_{\Delta} = \frac{  (m_j \hat{y}_{j}- m_i \hat{y}_{i} ) \sqrt{2 \rho_{\phi}}}{  \Delta m_{j i}^2 m_{\phi} } ~~~(i<j).
\end{equation}
Then, we can have a modification to the neutrino conversion
probability due to the shift in the neutrino mass diagonal terms,
\begin{equation}
    P_{\mu e} \simeq  \sin^2 2\theta \sin^2\Big[\frac{\Delta m^2 L}{4 E_{\nu}}\big(1+2\eta_{\Delta} \sin (m_{\phi } t)\big)  \Big].
\end{equation}

In the next section, we will implement the simulation of the effects of scalar field diagonal couplings in neutrino oscillations. 

\section{Simulation}
\label{simulation}
The ESS linac is projected to be fully
operational at 5 MW average power with an expected 2.5 GeV proton beam
currently under construction in Lund, Sweden. It will be an essential
user facility providing slow neutrons for research laboratories and
the industry. More importantly, for this study is the ESS$\nu$SB
initiative. A neutrino
super-beam facility that will benefit from the ESS production of
neutrons to search for the leptonic Dirac $CP$-violating phase
$\delta$~\cite{ESSnuSB:2013dql,ESSnuSB:2021azq,Alekou:2022emd}; the data taking it is
planned to start by 2035. It will investigate neutrino
oscillations around the second oscillation maximum with two baselines
in consideration at either 360 km or 540 km from the source. In
addition to measuring the leptonic Dirac $CP$-violating phase, the
ESS$\nu$SB facility may be employed to detect cosmological neutrinos
and neutrinos from supernova events and measure the proton lifetime.

This section presents the characteristics and assumptions performed in
our study. We use GLoBES~\cite{Huber:2004ka,Huber:2007ji} to simulate
an ESS$\nu$SB-like experiment with a 538 kt water Cherenkov
detector~\cite{MEMPHYS:2012bzz}. The information on the neutrino
fluxes is taken from Fig.~3 of the original
proposal~\cite{ESSnuSB:2013dql}, which corresponds to a 2.0 GeV proton
beam with $2.7\times10^{23}$ protons on target per year~\footnote{The
  annual operation period will be 208 days.} fixed at
5MW. Furthermore, the neutrino fluxes have been properly rescaled to
the corresponding baseline at $L =360$~km, or $L =540$~km distance, as
well as renormalized to the more recent simulation with 2.5 GeV
proton kinetic energy~\cite{ESSnuSB:2021azq}. The cross sections and
efficiencies in the detector follow the specifications from
Ref.~\cite{Campagne:2006yx}.  We assume an energy resolution which
follows a Gaussian distribution, with a width of
$\sigma(E)=12\%/\sqrt{E [\mbox{GeV}]}$ for electrons and
$\sigma(E)=10\%/\sqrt{E [\mbox{GeV}]}$ for muons, respectively. A
total of 12 bins uniformly distributed in the energy interval of
0.1-1.3 GeV were considered. Moreover, a 10-year exposure on a
far detector is considered in the form of $5$ years in neutrino mode
and $5$ years in antineutrino mode. Nevertheless, in our calibration
of the expected number of signal and background events, we have
assumed a one-year exposure to match the results from the updated
analysis released by the ESS$\nu$SB
collaboration~\cite{ESSnuSB:2021azq}. Unless otherwise specified, the
systematic errors are implemented as 10$\%$ signal normalization error
and 15$\%$ background normalization error for both appearance and
disappearance channels. In
Refs.~\cite{ESSnuSB:2013dql,ESSnuSB:2021azq}, the systematic errors
have been considered to be $5\%$ ($10\%$) for signal (background),
respectively. Ours are more conservative. Furthermore, a $0.01\%$
energy calibration error has been adopted for both types of
events. Our event rates reasonably reproduce~\footnote{Lately, the
  conceptual design report (CDR) for the ESS$\nu$SB experiment was
  released \cite{Alekou:2022emd}. From Table 8.1 of the CDR, an
  $\mathcal{O}(10\%)$ improvement on the expected background events
  with respect to our simulation was demonstrated. Signal events
  remain in good agreement, as shown in Table~[\ref{tab:1}]. As a
  result of our conservative assumptions, we do not expect
  considerable differences in our analysis.} the events reported in
Tables~2 and~3 of Ref.~\cite{ESSnuSB:2021azq}.

\begin{figure}[H]
		\begin{subfigure}[b]{0.49 \textwidth}
			\caption{  }
			\label{fig1}
			\includegraphics[width=\textwidth]{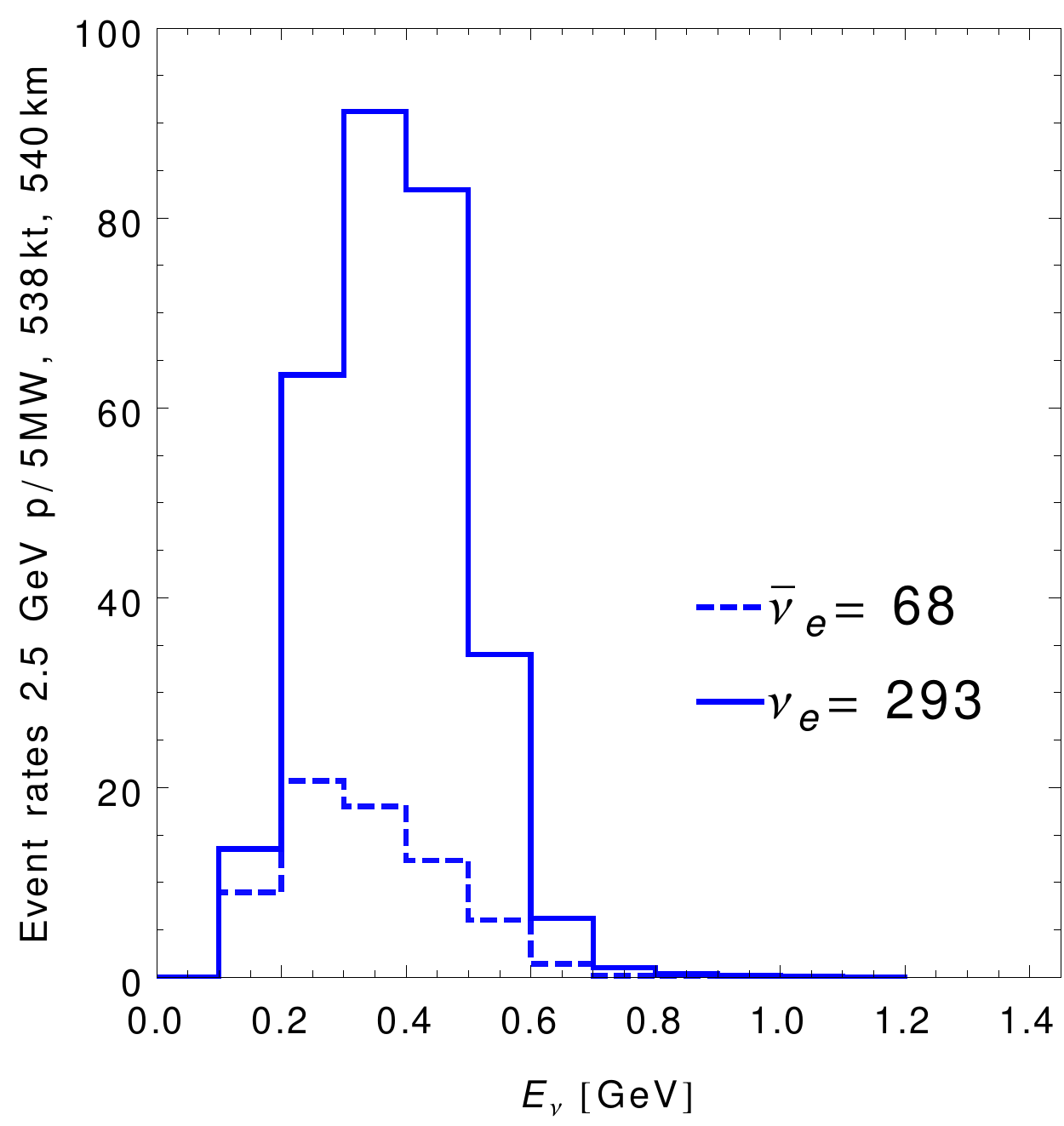}
		\end{subfigure}
		\hfill
		\begin{subfigure}[b]{0.49 \textwidth}
			\caption{}
			\label{fig2}
			\includegraphics[width=\textwidth]{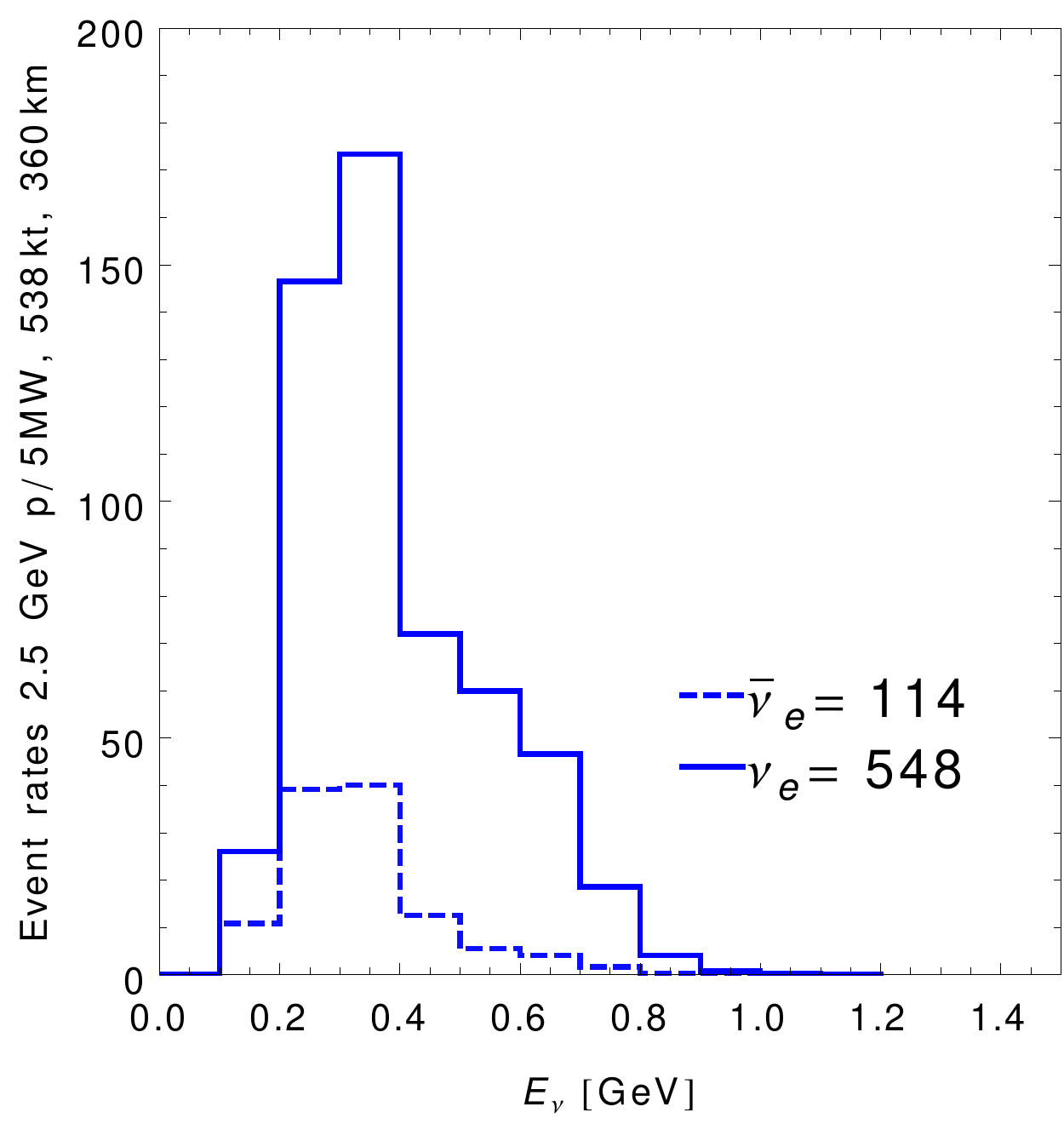}
		\end{subfigure}
		\hfill	
 \caption{Expected electron-neutrino and antineutrino appearance
   ($\nu_{e}/\bar{\nu}_e$) event rates per year for an ESS$\nu$SB
   setup using the neutrino oscillation parameters as given in
   \cite{ESSnuSB:2021azq}, the left panel displays the option of
   placing the detector at a baseline distance of $L=$540 km and the
   right panel for the $L=$360 km baseline case.  }
  \label{figg1}
\end{figure}

In Fig.~\ref{figg1}, we display our expected signal events as a
function of the neutrino energy, assuming one year of exposure for
both neutrinos and antineutrinos. The left panel shows the case of a
detector at a 540 km baseline from the source, and the right panel
considers a 360 km baseline.

\begin{table}[h]
\caption{\label{tab:1} Our expected electron-neutrino appearance signal and background events per year for an ESS$\nu$SB setup using the neutrino oscillation parameters as given in \cite{ESSnuSB:2021azq}.}
\centering
\begin{tabular}{| c | c | c | c | c | c | c |}
\hline \hline

Baseline run &$\nu_{e} (\Bar{\nu}_{e})$ sig. & $\nu_{\mu}(\Bar{\nu}_{\mu})$ misidentified ID&$\nu_{e}$ beam &  $\Bar{\nu}_{e}$ beam & NC bckg. & $\Bar{\nu}_{\mu}(\nu_{\mu}) \rightarrow \Bar{\nu}_{e}(\nu_{e})$ \\
\hline 

360 km $\nu$           & 548 & 87 & 164 & 0.2 & 37 & 3 \\
$~~~~~~~~~~~(\Bar{\nu}$) & 114 & 19 & 3  & 26  & 5 & 9 \\
540 km $\nu$           & 293 & 30 & 78 & 1 & 20 & 2 \\
$~~~~~~~~~~~(\Bar{\nu}$) & 68 & 6 &  1 & 12  & 4  & 6 \\ 

\hline \hline
\end{tabular}
\end{table} 

In Table~\ref{tab:1}, our total number of events per year for both
signal and background in the electron neutrino appearance channel are
presented accordingly for neutrinos and antineutrinos. We have
verified that with the inclusion of both electron and muon neutrino
data sets (Tables~\ref{tab:1} and \ref{tab:2}), our simulation
accurately replicates the precision studies on the atmospheric mixing
angle $\theta_{23}$ and mass squared splitting $\Delta m_{31}^2$ from
Fig.~8 of the ESS$\nu$SB Collaboration analysis
\cite{ESSnuSB:2021azq}. Once we have established our experimental
simulation and calibration of the number of events, in the following
section, we proceed to describe the characteristics of our study.
\section{Modulations from average distorted neutrino oscillations}
\label{mod}

In this section, we consider modulations of the ultralight scalar field in the regime $ \tau_{\nu} \ll \tau_{\phi} \ll
\tau_{\text{exp}}$ (average distorted neutrino oscillations). Here, $[
  \tau_{\nu} = (L/c) ]$ is the neutrino time of flight, $[\tau_{\phi}
  = 0.41 \times (10^{-14} ~\text{eV} / m_{\phi})$ seconds]
\cite{Dev:2020kgz} is the characteristic modulation period of the
scalar field and $[\tau_{\text{exp}}]$ is the lifetime of the
experiment. Under these circumstances, the oscillation effects from
mixing angles or mass splittings are too quick to be observed, but an
averaging effect on oscillation probabilities can be searched for
\cite{Berlin:2016woy,Krnjaic:2017zlz,Dev:2020kgz,Losada:2021bxx}.

Regarding the scalar field mass sensitivities at the
ESS$\nu$SB facility, for the baseline choices, $L=540$ km and $L=360$
km, the corresponding neutrino times of flight are
$\tau_{\nu}^{540}\approx 1.8 \times 10^{-3}$ sec and
$\tau_{\nu}^{360}\approx 1.2 \times 10^{-3}$ sec, respectively. The
total exposure of the experiment is $\tau_{\text{exp}}=10$
years. Therefore, for a ULSF modulation period $\tau_{\phi} \approx $
1 year, we expect an ultralight scalar field  mass sensitivity between
$2.0 \times 10^{-23}~\text{eV} \leq m_{\phi}^{540} \leq 1.3 \times
10^{-14}~\text{eV}$ and $2.0 \times 10^{-23}~\text{eV} \leq
m_{\phi}^{360} \leq 8.5 \times 10^{-15}~\text{eV}$, respectively.

The original proposal of the ESS$\nu$SB experiment
\cite{ESSnuSB:2013dql} is to search and optimize the physics potential
of leptonic Dirac $CP$-violating phase $\delta$ around the second
oscillation maximum. For this purpose, the electron neutrino
appearance channel is the natural choice for performing the
study. Despite its advantages for $CP$-violating phase searches, working
with this channel in the second oscillation maximum implies limited
statistics~\cite{Blennow:2019bvl}. Therefore, in our case, the
electron neutrino appearance channel might not be the optimal choice
to hunt for ultralight scalar field. Thus we will perform a combined
appearance and disappearance sensitivity study. In what follows, we
will describe the central features of this analysis and its results.

\subsection{Electron-neutrino appearance channel}
\label{enu}
In this part of the analysis, we describe the phenomenology to search
for a scalar field interaction with neutrinos via the ULSF parameter $\eta$ in the electron neutrino appearance channel. We
consider two proposed baseline distances, namely $L=$ 360 km and $L=$
540 km corresponding to the upcoming ESS$\nu$SB facility located in
Sweden, under the regime of average distorted neutrino oscillations
from the atmospheric mass squared difference $\Delta m_{31}^2$. Hence,
the effect of the ULSF interaction in the $P_{\mu e}$
oscillation probability for the two flavor approximation in vacuum is
given by
\begin{equation}
    P_{\mu e} \simeq \sin^2(2\theta_{23}) \Big \langle \sin^2 \big(  \frac{\Delta \Tilde{m}^2_{31} L}{4 E_{\nu}} \big) \Big \rangle,
\end{equation}
with $\Delta \Tilde{m}_{31}^2 =$ $\Delta m_{31}^2 \big[1+2\eta \sin(m_{\phi}t) \big]$; the average over the mass squared difference is given by~\cite{Krnjaic:2017zlz,Dev:2020kgz,Losada:2021bxx}
\begin{equation}
\label{eq3}
\begin{split}
     ~~~~~~~~~~~~~~~~~~~~~~\Big \langle \sin^2 \big(  \frac{\Delta \Tilde{m}^2_{31} L}{4 E_{\nu}} \big) \Big \rangle &= \frac{1}{\tau_{\phi}}\int^{\tau_{\phi}}_{0} dt \sin^2 \big[ \Delta_{31}\big(1+2 \eta \sin (m_{\phi} t)\big)\big] \\
      &\simeq \sin^2(\Delta_{31}) + 2\cos(2 \Delta_{31}) \Delta_{31}^2\eta^2+\mathcal{O}\big(\Delta_{31}^2\eta^2\big)^2, ~~~~~~~~~~~~~~~~~~~~~~~~
\end{split}
\end{equation}
where $\tau_{\phi}$ is the ULSF period and
$\Delta_{31} = \frac{\Delta m^2_{31} L}{4 E_{\nu}}$.  More generally,
the oscillation probability in the appearance channel considering
matter effects used in this study follows from \cite{Nunokawa:2007qh}:
\begin{equation}
\label{prob}
\begin{split}
    P(\nu_{\mu} \rightarrow \nu_{e}) & = \sin^2\theta_{23} \sin^2 2 \theta_{13} \frac{\sin^2(\Delta_{31}-aL)}{(\Delta_{31}-aL)^2} \Delta_{31}^2   \\
   &+ \sin 2 \theta_{23} \sin 2 \theta_{13} \sin 2 \theta_{12} \frac{\sin(\Delta_{31}-aL)}{(\Delta_{31}-aL)} \Delta_{31} \frac{\sin(aL)}{(aL)} \Delta_{21}\cos(\Delta_{31}+\delta) \\
   & + \cos^2\theta_{23} \sin^2 2 \theta_{12} \frac{\sin^2(aL)}{(aL)^2} \Delta_{21}^2,
\end{split}
\end{equation}
where $a = G_{F}N_{e}/\sqrt{2}$ is the matter potential, $G_{F}$ is
the Fermi constant, $N_{e}$ is the electron density, and $\Delta_{ij}
= \Delta m_{ij}^2 L/ 4E_{\nu}$. For antineutrinos, we replace $a
\rightarrow -a$ and $\delta \rightarrow -\delta$.  Furthermore, in
this channel, the scalar field parameter, $\eta$, appears in the form
of an average over the frequency of oscillations uniquely mediated by
the atmospheric mass squared splitting $\Delta m_{31}^2$ as shown in
Eq.~(\ref{eq3}). The details on the expected signal and background
events for this channel were discussed in the previous section
Sec.~\ref{simulation}. As far as neutrino oscillation parameters are
concerned, the true values used in this analysis are $\Delta
m^{2}_{21}=7.5 \times 10^{-5}~\text{eV}^{2}$, $\Delta m^{2}_{31} =
2.55 \times 10^{-3}~\text{eV}^2$, $\theta_{12} = 34.3^{\circ}$,
$\theta_{13} = 8.53^{\circ}$, $\theta_{23} = 49.26^{\circ}$, and
$\delta = 194^{\circ}$, corresponding to the best-fit values for
normal ordering from Salas \emph{et
  al.}~\cite{deSalas:2020pgw}. \footnote{We have verified that our
  results do not significantly change by using the best-fit values
  from Ref.~\cite{Esteban:2020cvm}.} For oscillation parameter
priors, we assume a 1$\sigma$ error of $ 5\%$ for $\Delta m^{2}_{21}$,
$\Delta m^{2}_{31}$, $\theta_{12}$, and $\theta_{23}$. We also assume
$3\%$ for $\theta_{13}$ and $ 10\%$ for the leptonic $CP$-violating
phase $\delta$~\cite{ESSnuSB:2013dql}. In addition, matter effects
were considered, for both baselines, with a constant density of $\rho
= 2.8~\text{g/cm}^3$~\cite{Dev:2020kgz}.
\subsection{Muon-neutrino disappearance channel}
Here we shift our attention to the muon-neutrino disappearance
channel, which benefits from a larger event signal with minimal
background contamination \cite{ESSnuSB:2021azq},\footnote{Regarding
  the muon-neutrino sample, a detailed physics reach from the muon
  disappearance channel is not presented in the ESS$\nu$SB CDR
  \cite{Alekou:2022emd}, our simulation is based on the results from
  Sec.~3.2 of \cite{ESSnuSB:2021azq}.} as displayed in
Table~\ref{tab:2}. Besides the agreement on the expected signal and
background events, we verify that our simulation accurately replicates
the precision studies on the atmospheric mixing angle $\theta_{23}$
and mass squared splitting $\Delta m_{31}^2$ from Fig.~8 of the
ESS$\nu$SB Collaboration analysis \cite{ESSnuSB:2021azq}. The muon
-neutrino disappearance channel is optimal for investigating the ULSF oscillation phenomenology. We include in our
analysis the corresponding matter effects needed for a complete study
of the ESS case~\cite{Agarwalla:2014tpa}, although it does not
significantly improve the sensitivity to $\Delta m_{31}^2$ in our
study, see e.g., the authors of
Refs.~\cite{Blennow:2019bvl,Agarwalla:2014tpa,Chakraborty:2019jlv}. In
the standard oscillation case, neglecting $\Delta m^2_{21}$ effects,
the survival probability will be given by~\cite{Choubey:2005zy}
\begin{equation}
\label{prob2}
\begin{split}
    P(\nu_{\mu} \rightarrow \nu_{\mu}) & \approx 1- \sin^2\theta_{13}^{\text{M}} \sin^2 2\theta_{23} \sin^2\Big[\frac{1}{2}(\Delta_{31}-\Delta_{31}^{\text{M}}+\Delta_{A})\Big] \\
   &~~~~~~ -\cos^2\theta_{13}^{\text{M}} \sin^2 2\theta_{23} \sin^2\Big[\frac{1}{2}(\Delta_{31}+\Delta_{31}^{\text{M}}+\Delta_{A})\Big] \\
   &~~~~~~ -\sin^2 2 \theta_{13}^{\text{M}} \sin^4 \theta_{23} \sin^2\Delta_{31}^{\text{M}},
\end{split}
\end{equation}
where $\Delta_{31}^{\text{M}}=(\Delta m^{2}_{31})^{\text{M}}
L/4E_{\nu}$, $\Delta_{A}=A L/4E_{\nu}$ being $\theta_{13}^{\text{M}}$
and ($\Delta m^{2}_{31})^{\text{M}}$, the effective reactor mixing
angle $\theta_{13}$ and atmospheric mass-squared difference $\Delta
m_{31}^2$ in matter, with $A$ given by $A(\text{eV}^2)=0.76 \times
10^{-4} \rho (\text{g/cm}^{3}) E_{\nu} (\text{GeV})$. For
antineutrinos, we replace $A \rightarrow -A$. As we have already
discussed, the scalar-field interaction parameter, $\eta$, enters in
as an average over the oscillation frequency mediated by the
atmospheric mass-squared splitting $\Delta m_{31}^2$ as shown in
Eq.~(\ref{eq3}), in a similar way as in the electron neutrino
appearance channel. For the relevant neutrino oscillation parameters
in this channel (including matter effects), we follow the same
specifications as the electron neutrino appearance channel.
\begin{figure}[H]
		\begin{subfigure}[b]{0.49 \textwidth}
			\caption{ }
			\label{figa}
			\includegraphics[width=\textwidth]{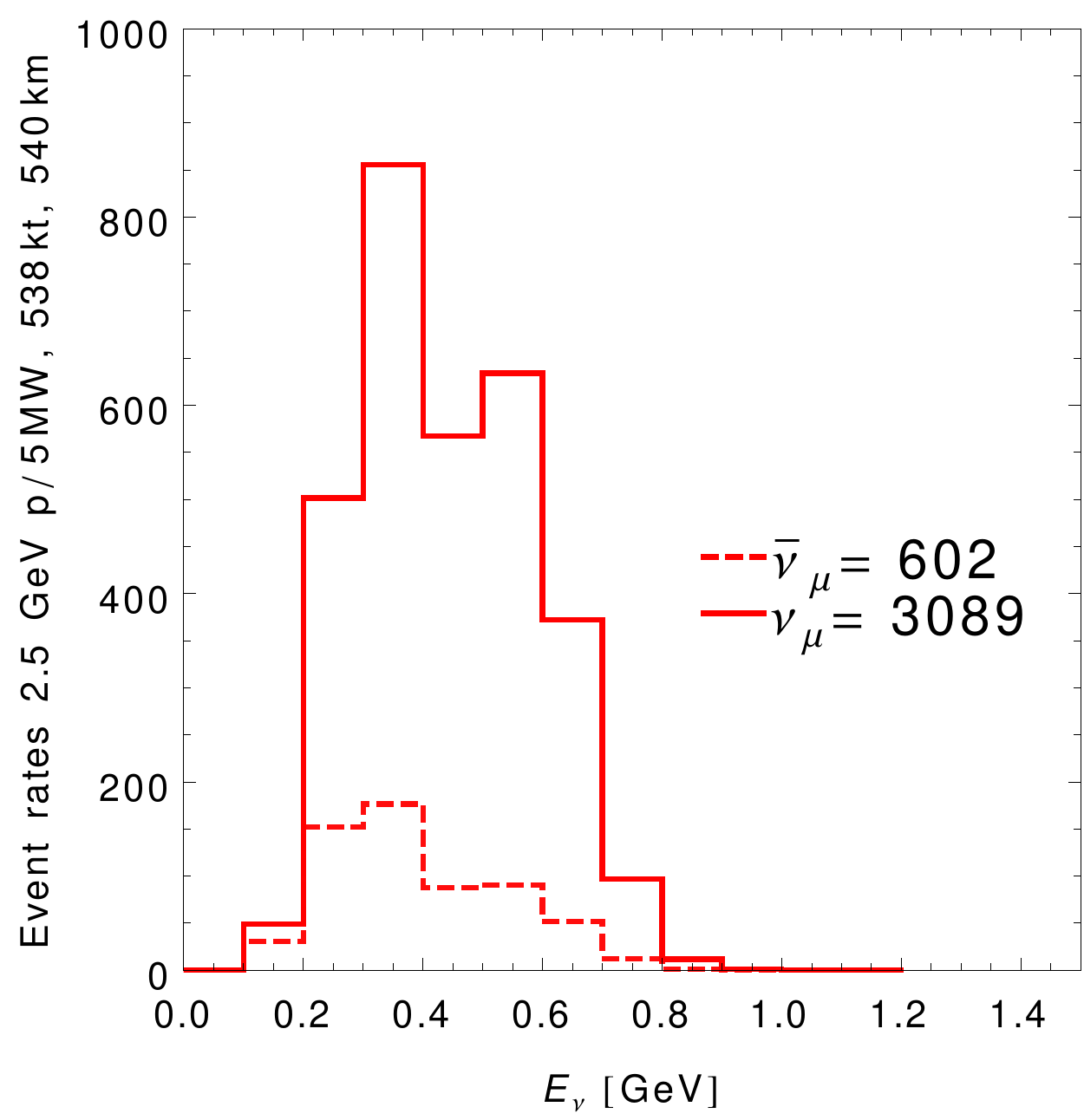}
		\end{subfigure}
		\hfill
		\begin{subfigure}[b]{0.49 \textwidth}
			\caption{}
			\label{figb}
			\includegraphics[width=\textwidth]{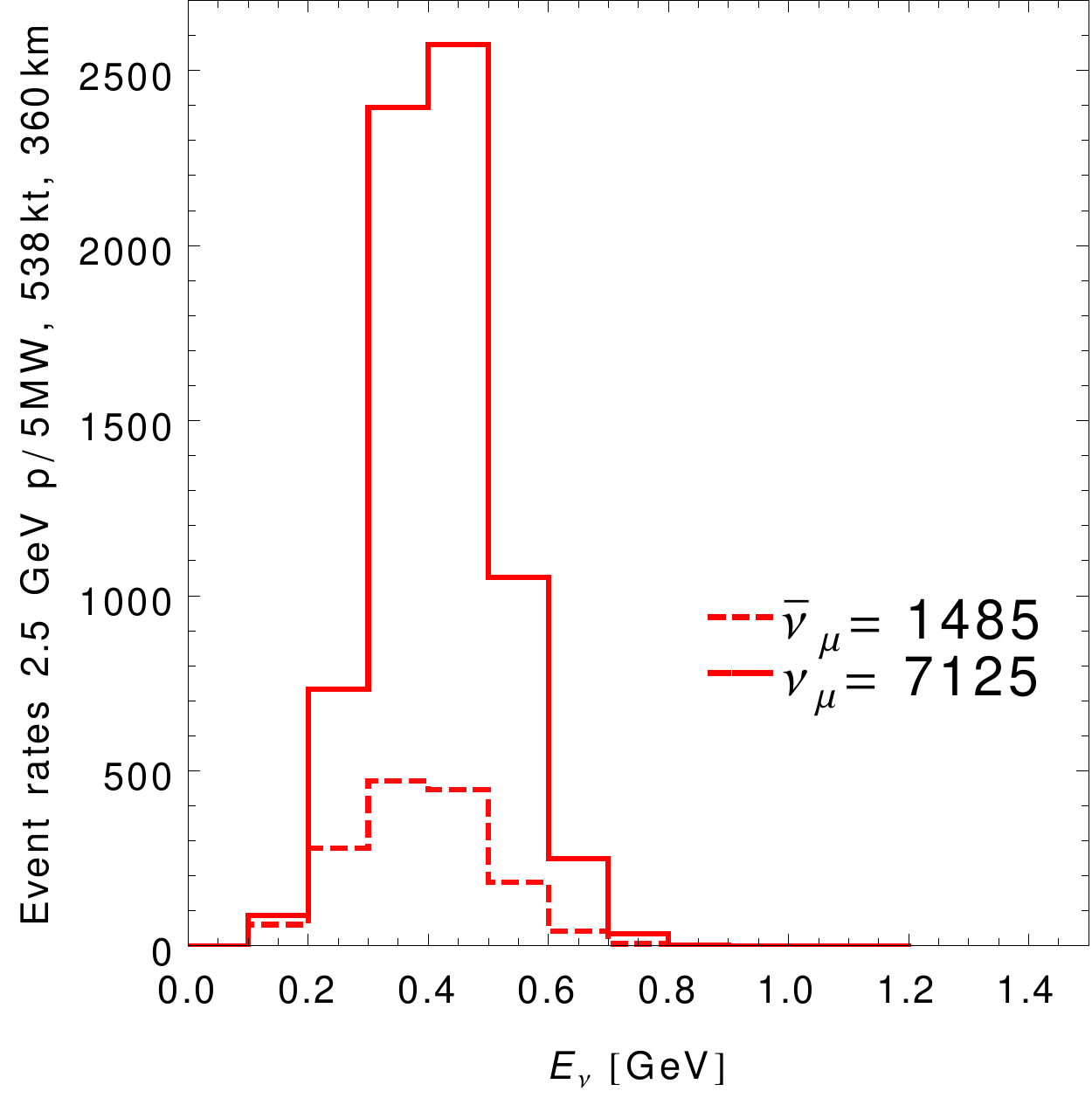}
		\end{subfigure}
		\hfill	
 \caption{Expected muon-neutrino and antineutrino disappearance
   ($\nu_{\mu}/\bar{\nu}_{\mu}$) event rates per year for an
   ESS$\nu$SB setup using the neutrino oscillation parameters as given
   in \cite{ESSnuSB:2021azq}, the left panel displays the baseline
   option of placing the detector at a distance of $L =$540 km and the
   right panel for $L =$360 km baseline case.}
  \label{figaa}
\end{figure}
In Fig.~\ref{figaa}, we display our expected signal events as a
function of the neutrino energy, assuming one year of exposure, for
both neutrinos and anti-neutrinos. The left panel is for the option of
placing the detector at a baseline of 540 km from the source, and the
right panel represents the baseline option at 360 km from the source.

\begin{table}[h]
\caption{\label{tab:2} Our expected muon-neutrino disappearance signal
  and background events per year for an ESS$\nu$SB setup using the
  neutrino oscillation parameters as given in \cite{ESSnuSB:2021azq}.}
\centering
\begin{tabular}{| c | c | c | c | }
\hline \hline

Baseline run &$\nu_{\mu} (\Bar{\nu}_{\mu})$ sig. & $\Bar{\nu}_{\mu} (\nu_{\mu}) \rightarrow \Bar{\nu}_{\mu}(\nu_{\mu}) $  & NC bckg. \\
\hline 

360 km $\nu$           & 7125 & 54 & 87  \\
$~~~~~~~~~~~(\Bar{\nu}$) & 1485 & 132 & 13    \\
540 km $\nu$           & 3089 & 27 & 38    \\
$~~~~~~~~~~~(\Bar{\nu}$) & 602  & 67  &  6  \\ 

\hline \hline
\end{tabular}
\end{table} 

In Table~\ref{tab:2}, the total number of events per year for signal
and background in the muon-neutrino disappearance channel is
introduced for both neutrinos and antineutrinos. In our analysis, we
consider only the $\Bar{\nu}_{\mu} (\nu_{\mu}) \rightarrow
\Bar{\nu}_{\mu}(\nu_{\mu})$ and $\nu_{\mu} (\bar{\nu}_{\mu})$ neutral
current backgrounds for positive (negative) polarity, respectively,
since they are the main contributions to this channel. As a result,
there is essentially no background interference in the muon-neutrino
disappearance channel.
 
\subsection{Scalar-field sensitivity}
\label{sfdm}
In this subsection, we introduce our results of the ULSF searches via neutrino oscillations at the ESS$\nu$SB from the
combined analysis at both appearance and disappearance channels.  We
have already stated that the disappearance channel will give a more
restrictive result for this kind of search due to its higher
statistics. Still, we include in our analysis both the electron
appearance as well as the muon disappearance channels, to be the most
sensitive as possible.

We employ a chi-squared test to quantify the statistical significance
of the ULSF sensitivity, which is given by the
adding the two channels using both neutrino and anti-neutrino data
sets. The $\chi^2$ function\footnote{More details on the
  implementation of the $\chi^2$ function, systematical errors and
  priors in the GLoBES software~\cite{Huber:2004ka,Huber:2007ji} can
  be found in \cite{Huber:2002mx}.} is given as
\begin{equation}
    \chi^2 = \sum_{\ell} \tilde{\chi}^2_{\ell} +\chi^2_{\text{prior}},
\end{equation}
where the corresponding $\tilde{\chi}^2_{\ell}$ function for each
channel $\ell=\{\nu_{\mu}(\Bar{\nu}_{\mu})\rightarrow \nu_{e}
(\Bar{\nu}_{e}),\nu_{\mu}(\Bar{\nu}_{\mu})\rightarrow \nu_{\mu}
(\Bar{\nu}_{\mu}) \}$ in the large data size limit is given by
\begin{equation}
    \tilde{\chi}^2_{\ell} = \min_{\xi_{j}} \Bigg[ \sum_{i}^{n_{\text{bins}}}\frac{\big(N_{i, \text{true}}^{3\nu}-N_{i, \text{test}}^{3 \nu+\eta}( \Omega, \eta, \{\xi_{j}\}) \big)^2}{\sigma_{i,\text{true}}^2}+\sum_{j}^{n_{\text{syst.}}} \Big(\frac{\xi_{j}}{\sigma_{j}}\Big)^2 \Bigg].
\end{equation}
The $N_{i}^{3\nu}$ are the simulated events at the $i$th energy bin
considering the standard three neutrino oscillations as a true
hypothesis. $N_{i}^{3\nu +\eta}$ are the computed events at the $i$th
energy bin with the model assuming scalar field 
oscillations. $\Omega = \{\theta_{12}, \theta_{13}, \theta_{23},
\delta, \Delta m_{21}^2, \Delta m^2_{31}\}$ is the set of oscillation
parameters, $\eta$ is the ULSF parameter and $\{\xi_{j}\}$ are the
nuisance parameters to account for the signal, background
normalization, and energy calibration systematics
respectively. Moreover, $\sigma_{i} = \sqrt{N_{i}^{3\nu}}$ is the
statistical error in each energy bin while $\sigma_{j}$ are the
signal, background normalization, and energy calibration errors (see
Sec.~\ref{simulation}). Furthermore, implementation of external input
for the standard oscillation parameters on the $\chi^2$ function is
performed via Gaussian priors
\begin{equation}
    \chi^2_{\text{prior}}=  \sum_{k}^{n_{\text{priors}}} \frac{\big(\Omega_{k,\text{true}}-\Omega_{k,\text{test}}\big)^2}{\sigma^2_{k}}, 
\end{equation}
the central values of the priors $\Omega_{k}$ are set to their true or
best-fit value for normal ordering
\cite{deSalas:2020pgw}. $\sigma_{k}$ is the uncertainty on the
oscillation prior, which corresponds to a 1$\sigma$ error of $5\%$ for
$\Delta m^{2}_{21}$, $\Delta m^{2}_{31}$, $\theta_{12}$, and
$\theta_{23}$, $3\%$ for $\theta_{13}$, and $ 10\%$ for the leptonic
$CP$-violating phase $\delta$~\cite{ESSnuSB:2013dql}, the summation
index $k$ runs over the corresponding test oscillation parameters to
be marginalized.

Besides, the expected number of events at the $i$th energy bin is
calculated as \cite{Huber:2002mx}
\begin{equation}
\label{events}
    N_{i}=\frac{\mathcal{N}}{L^2} \int^{E_{i}^{\prime}+ \Delta E_{i}^{\prime}/2}_{E^{\prime}_{i}-\Delta E_{i}^{\prime}/2} d E^{\prime} \int^{\infty}_{0} dE~\Phi_{\nu}(E)  \sigma_{\nu}(E) P_{\alpha \beta}(E) K(E,E^{\prime})\varepsilon(E^{\prime}),
\end{equation}
where $E$ is the true neutrino energy, $E^{\prime}$ is the
reconstructed neutrino energy, $\Delta E_{i}^{\prime}$ is the bin size
of the $i$th energy bin, $\mathcal{N}$ is a constant normalization
factor that accounts for the mass-year exposure and beam power, $L$
is the baseline distance, $\Phi_{\nu}(E)$ is the energy-dependent
neutrino flux, $\sigma_{\nu}(E)$ is the energy-dependent cross
section, $P_{\alpha \beta}(E)$ is the neutrino-oscillation
probability, $K(E,E^{\prime})$ is the energy response-model or energy-resolution of the experiment, and $\varepsilon(E^{\prime})$ is the
energy-dependent efficiency. Moreover, the neutrino-oscillation
probability $P_{\alpha \beta}(E)=P_{\nu_{\alpha} \rightarrow
  \nu_{\beta}}(E,L,\rho,\theta_{12}, \theta_{23},\theta_{13},\Delta
m_{31}^2, \Delta m_{21}^2, \delta)$; the energy-resolution function,
which relates the true and reconstructed neutrino energies follow a
Gaussian distribution
\begin{equation}
    K(E,E^{\prime})= \frac{1}{\sqrt{2 \pi} \sigma_{R}(E)} \exp \Bigg\{-\frac{(E-E^{\prime})^2}{2 \sigma_{R}^2(E)} \Bigg\},
\end{equation}
where $\sigma_{R}(E)= \beta \sqrt{E/\text{GeV}} ~\text{GeV} $, in our
case we assume $\beta=(0.12,0.10)$ for ($e^{-},\mu^{-}$) respectively.

Furthermore, $N_{i}^{3\nu} = S^{3\nu}_{i}+B^{3\nu}_{i}$, $S_{i}^{3
  \nu}$, and $B_{i}^{3\nu}$ are the simulated signal and background
events in each energy bin within the standard three neutrino
oscillations framework, as described in Tables~\ref{tab:1}, and
\ref{tab:2}. They were computed according to Eq.~(\ref{events})
with true oscillation parameters fixed to their best-fit point for
normal ordering (see Sec.~\ref{enu}). In addition, $N_{i}^{3\nu +
  \eta} = S^{3\nu+ \eta}_{i}\big(1+
\xi_{1}+g(E^{\prime})\xi_{3}\big)+B^{3\nu + \eta}_{i}\big(1+
\xi_{2}+g(E^{\prime})\xi_{4}\big)$, $S_{i}^{3 \nu +\eta}$, and
$B_{i}^{3\nu+\eta}$ are the corresponding signal and background events
in each energy bin, assuming the model with scalar field 
oscillations where $P_{\alpha \beta}(E) \rightarrow P_{\alpha
  \beta}(E, \eta)$. Likewise, the $\{\xi_{j}\}$ are the nuisance
parameters describing the systematic errors, where $\xi_{1}$ and
$\xi_{2}$ account for the signal and background normalization,
respectively, whereas $\xi_{3}$ and $\xi_{4}$ account for the signal
and background energy calibration, the energy calibration function
$g(E^{\prime})$ is
\begin{equation}
   g(E^{\prime}) = \frac{(E_{i}^{\prime}-\bar{E}^{\prime})}{(E^{\prime}_{max}-E^{\prime}_{min})},
\end{equation}
where $E_{i}^{\prime}$ is the mean reconstructed energy at the $i$th
energy bin; $\bar{E}^{\prime} = \frac{1}{2} (E^{\prime}_{max} +
E^{\prime}_{min})$ is the median of the energy interval,
$E^{\prime}_{min} = 0.1$~GeV is the minimum energy of the
reconstructed energy window while $E^{\prime}_{max} =1.3$~GeV is the
maximum energy of the reconstructed energy window.

The sensitivity contours were computed based on the $\Delta \chi^2 =
\chi^2 - \chi ^2_{\text{min}}$ distribution, scanning over the test
parameter pairs, either ($\sin^2 \theta_{23}$, $\eta$) or ($\Delta
m^2_{31}$, $\eta$), with the ULSF parameter $\eta$ arising from the
average modulation on the atmospheric mass-squared splitting $\Delta
\Tilde{m}_{31}^2$, as given in Eq.~(\ref{eq3}). All the remaining test
oscillation parameters are marginalized over. The standard three
neutrino oscillation picture is assumed as true hypothesis; the
boundary of the corresponding allowed regions were determined by
mapping the $\Delta \chi^2$ to corresponding confidence levels using a
$\chi^2$-distribution, assuming Wilks theorem for two degrees of
freedom.

\begin{figure}[H]
		\begin{subfigure}[b]{0.47 \textwidth}
			\caption{  }
			\label{fig33}
			\includegraphics[width=\textwidth]{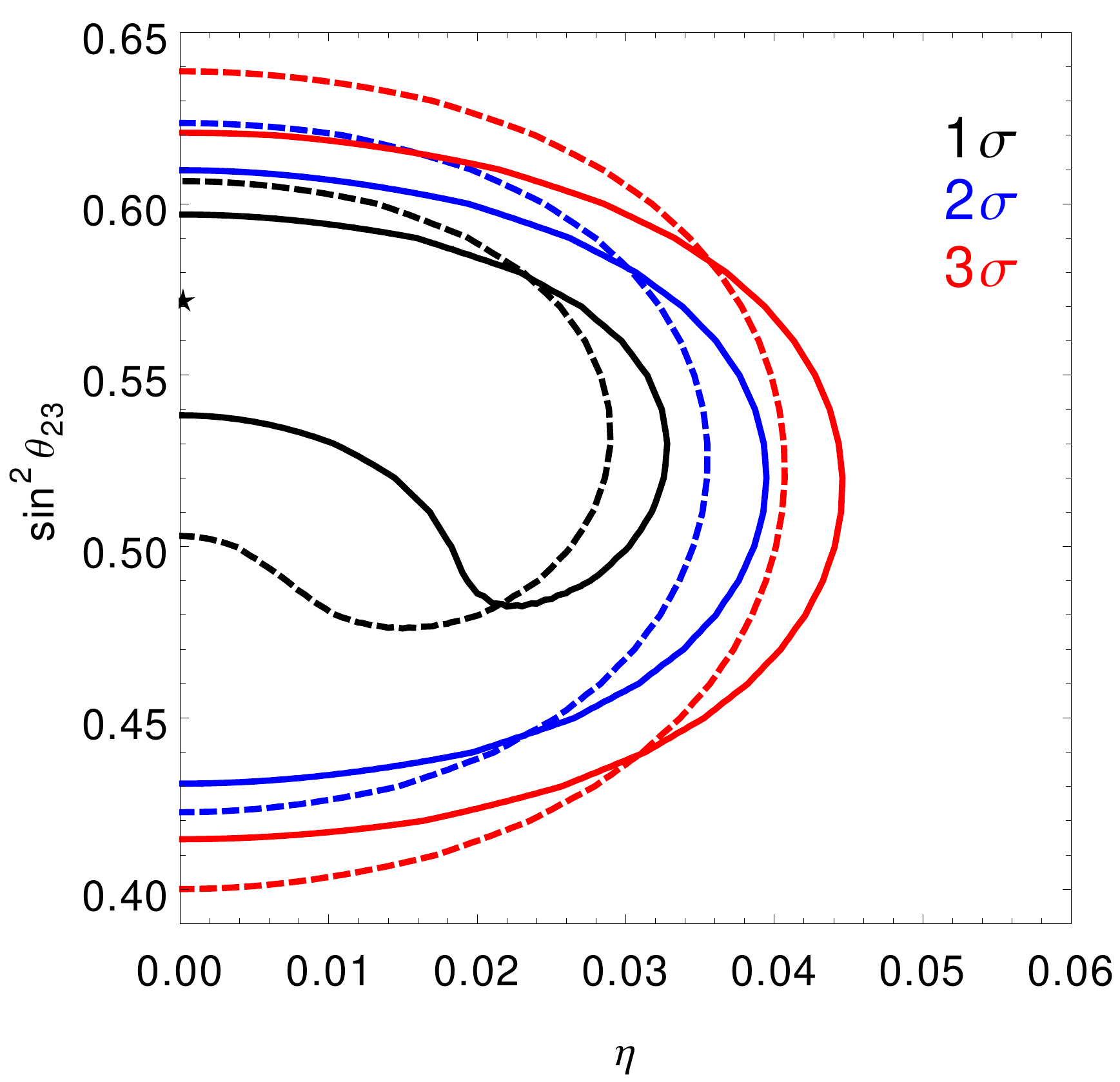}
		\end{subfigure}
		\hfill
		\begin{subfigure}[b]{0.5 \textwidth}
			\caption{}
			\label{fig44}
			\includegraphics[width=\textwidth]{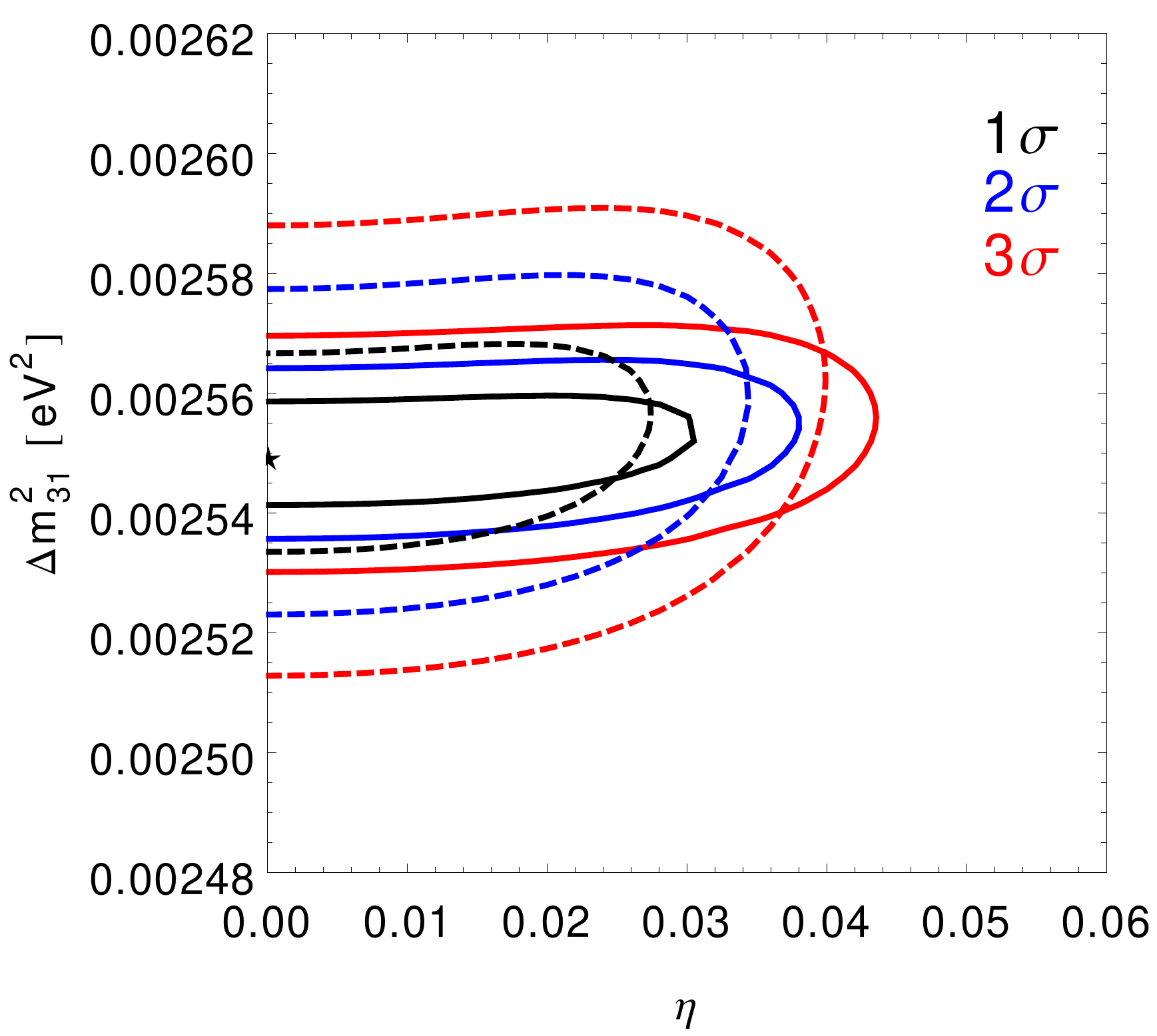}
		\end{subfigure}
		\hfill	
 \caption{Expected sensitivity to the scalar field 
   parameter $\eta$ from the atmospheric mass squared modulation at
   the ESS$\nu$SB setup; the baseline choice of $L = 540$ km is shown
   in dashed lines, whereas the $ L = 360$ km in solid lines. The
   contours inside the black, blue, and red lines are the
   sensitivities at $1\sigma$, $2\sigma$, and $3\sigma$, respectively.
 }
  \label{figj}
\end{figure}

In Fig.~\ref{figj}, we show our sensitivities to the scalar field
 scenario. The left panel displays the expected sensitivity
to the ULSF in the $(\eta, \sin^2
\theta_{23})$ plane, whereas the right panel shows the expected
sensitivity in the $(\eta, \Delta m^2_{31})$ plane, respectively. The
contours inside the black, blue, and red lines are the sensitivities
at ($1\sigma$, $2\sigma$, and $3\sigma$), which correspond to $\Delta
\chi^2 =$ (2.3, 6.18, and 11.83) accordingly. The star in the
$\eta-\sin^2 \theta_{23}$ and $\eta-\Delta m^2_{31}$ planes represent
the best-fit point used in the simulated data under the true
hypothesis. The solid lines assume an ESS$\nu$SB setup with a baseline
of $L= 360$~km, while the dashed lines represent a baseline of $L=
540$~km.

We observe that the impact of the normalization systematic error is
relevant for the $L = 540$ km baseline, mainly due to a decrease in
the signal events, spoiling the precision on the atmospheric mixing
parameters, $\theta_{23}$ and $\Delta m^2_{31}$. However, the
sensitivity to the scalar field parameter is not
considerably affected. As a result, we anticipate a ULSF parameter sensitivity of $\eta < 0.043 ~(0.039)$ at 3$\sigma$
and $\eta < 0.032 ~(0.028)$ at 90$\%$ confidence limit (C.L.) i.e., $\Delta
\chi^2=4.61$, at the $L=360$ km ($L =540$ km) baselines. Besides, the
allowed values at $3\sigma$ for the mixing angle are $0.41\lesssim
\sin^2 \theta_{23}\lesssim 0.62$, and for the atmospheric mass-squared
difference $ 2.53 \times 10^{-3}~\text{eV}^2 \lesssim \Delta
m^2_{31}\lesssim 2.57 \times 10^{-3}~\text{eV}^2$, at the $L=$ 360 km
baseline option, while at the $L=$ 540 km baseline, the allowed values
at $3\sigma$ are $0.40\lesssim \sin^2 \theta_{23}\lesssim 0.64$ and $
2.51 \times 10^{-3}~\text{eV}^2 \lesssim \Delta m^2_{31}\lesssim 2.59
\times 10^{-3}~\text{eV}^2$.

We can notice the relevance of the expected sensitivity found here by
comparing it with other studies for different types of neutrino
physics. For instance, for solar neutrino experiments, it has been
pointed out~\cite{Berlin:2016woy} that an order $\mathcal{O}(10\%)$
anomalous modulation on the neutrino fluxes, due to an ultralight
($m_{\phi} \sim 10^{-22}$ eV) scalar coupling with neutrinos could
happen at solar neutrino experiments. This result translates into a
bound to the scalar field parameter of $\eta \sim 0.1$. In
addition, a projected sensitivity of $\eta \sim 1.5 \%$ and $\eta \sim
7.5 \%$ via $\Delta m_{31}^2$ smearing are expected at the JUNO and DUNE experiments, respectively
\cite{Krnjaic:2017zlz}. Furthermore, bounds due to the modulations
from the mass-squared splitting $\Delta m_{31}^2$ have been recently
studied in Ref.~\cite{Losada:2021bxx}, reporting a 1$\sigma$
sensitivity to the $\eta$ parameter for different neutrino
experiments, such as Daya Bay; with a scalar field  bound
of $\eta \sim 0.08$ from the electron antineutrino disappearance
channel. Similarly, the JUNO experiment is projected to be sensitive
to $\eta \sim 0.005$, and from the electron neutrino appearance
channel, a $\eta \sim 0.01$ bound is expected for both DUNE and
Hyper-Kamiokande, respectively. Regarding DUNE, the authors of
Ref.~\cite{Dev:2020kgz} obtained a 1$\sigma$ sensitivity of $\eta \sim
0.035$. Hence, compared to our projected 1$\sigma$ sensitivities to
the ULSF parameter at the ESS$\nu$SB setup of $\eta \sim 0.03$ from
the $L= 360$ km baseline option and $\eta \sim 0.027$ from the $L=
540$ km baseline, competitive bounds on the ULSF parameter can be
achieved. Consequently, we expect that with the inclusion of the muon-neutrino disappearance data set, we can motivate to extend the main
physics program at the ESS$\nu$SB to search for this type of physics.

\subsection{Some cosmological implications of the neutrino-ULSF interaction}

Before finishing this section, we would like to discuss some important characteristics and details of the scalar field in cosmology --they can be relevant when interpreting our results.

We start this subsection by noticing, as stated in Ref~\cite{Berlin:2016woy}, that the $\hat{y} \phi$ contribution in Eq.~(\ref{int}) will induce quantum corrections\footnote{The tadpole contribution vanishes for off-diagonal couplings where $\text{Tr}[m_{\nu} \hat{y}_{ij}]=0$ in flavor space \cite{Losada:2021bxx}.} to the ULSF potential
\begin{equation}
    V_{\phi} = \frac{m_{\phi}^2 }{2}\phi^2  + \hat{y} n_{\nu} \phi + \mathcal{O}(\hat{y}^2) + \dots.
\end{equation}
This contribution will produce a shift of the $\phi$ vacuum expectation value (VEV) that goes as $\hat{y} n_{\nu} \phi$, where $n_{\nu}$ is the neutrino number density. This shift could jeopardize the features of the ULSF as dark matter since the VEV of $\phi$ will be displaced at $T \sim$~GeV, 
\begin{equation}
    \langle-\phi\rangle \sim \frac{\hat{y} n_{\nu} }{ m_{\phi}^2 }\sim 10^{21} ~\text{GeV}~ \Big( \frac{\hat{y}}{10^{-40}} \Big)  \Big( \frac{n_{\nu}} {0.1 \text{GeV}^3} \Big)
 \Big( \frac{10^{-22}\text{eV} }{m_{\phi}} \Big).
\end{equation}
For this reason, it is important to consider that the temperature associated with the phase transition that sets the initial conditions for the ULSF field it is well below $1$~MeV~\cite{Berlin:2016woy}.

If $\phi$ begins to oscillate at $T_{\text{osc}} \sim \sqrt{M_{\text{Pl}} m_{\phi}}~ (m_{\phi}/10^{-22}\text{eV}) \sim$ keV \cite{Hui:2016ltb}, around this temperature, the redshift $z_{\text{osc}}\sim 2 \times10^6$ and the average density $\rho_{\phi}(z) \sim 10^{-5} \rho_{\phi,\odot} (1+z)^3$. Thus, the scalar-field amplitude $\phi_{\text{keV}} (z= 2 \times10^6) \simeq 10^{26}$ eV $\simeq 10^{17}$ GeV. From the sensitivity to the ULSF parameter $\eta \sim$ 0.01 at ESS$\nu$SB, using Eq.~(\ref{eta}) we obtain $\hat{y} \sim 10^{-23} ~(m_{\phi}/ 10^{-22} \text{eV})$ at present time. However if the coupling $\hat{y}$ remains unmodified back to the oscillation temperature, the VEV of the scalar field ($ \langle- \phi \rangle = \phi_{\text{min}}$) is
    \begin{equation}
    \langle-\phi\rangle \sim \frac{\hat{y} n_{\nu} }{ m_{\phi}^2 }\sim 10^{20} ~\text{GeV}~ \Big( \frac{\hat{y}}{10^{-23}} \Big)  \Big( \frac{n_{\nu}} {10^{-19} \text{GeV}^3} \Big)
 \Big( \frac{10^{-22}\text{eV} }{m_{\phi}} \Big),
\end{equation} thus, $\phi_{\text{min}} \sim 10^{3}$ $\phi_{\text{keV}}$ (which requires a coupling $\hat{y} \lesssim 10^{-27} ~(m_{\phi}/ 10^{-22} \text{eV})$ to avoid large corrections to the neutrino mass, well below the sensitivity of any terrestrial neutrino oscillation experiment). Furthermore, the contributions to the neutrino mass are $\hat{y} \phi_{\text{keV}} \simeq 10^3$ eV and $\hat{y} \phi_{\text{min}} \simeq 10^6$ eV, hence the effective neutrino mass at this temperature would be $\Tilde{m}(t) \simeq 10^6$ eV.

On the other hand, around matter-radiation equality where $T_{\text{eq}} \sim 1$ eV; the potential $\hat{y} \phi \sim 10^{-23} \times 10^{22}$ eV $\sim 0.1$ eV and $\hat{y} (\phi + \phi_{\text{min}}) \sim 0.11$ eV. Besides, the effective neutrino mass $\Tilde{m}(t) \simeq m_{\nu} + \hat{y} (\phi + \phi_{\text{min}})$, assuming a bare neutrino mass $m_{\nu} \sim 0.1$ eV we obtain $\Tilde{m}(t) \simeq 0.21$ eV, which is within the limit of the CMB constraint $\sum_i m_i < 0.23$ eV \cite{Planck:2015fie}. Therefore, without invoking any  model realization, unless the temperature associated with the phase transition of the ULSF occurs after matter-radiation equality (say at $T \sim m_{\nu} \sim 0.1$ eV), strong constraints apply to the neutrino-ULDM scenario at ESS$\nu$SB. Moreover, if this transition happens after the formation of the CMB, the ULSF can only constitute a fraction of the DM~\cite{Krnjaic:2017zlz}. Therefore, in our analysis for the ESS$\nu$SB, we can consider the scalar field as being a component of the dark matter, or to be just an ultralight scalar field without participating of the DM components \cite{Capozzi:2018bps}.

In the case that the phase transition can happen at temperatures of the order of $10$~MeV, there could be potential contributions to the effective number of neutrino species $N_{\text{eff}}$ during big bang nucleosynthesis (BBN). In this scenario, the interaction term $\hat{y} \phi$ could increase $N_{\text{eff}}$ during BBN. However, the density of thermal neutrinos may have dominated the ULSF potential throughout this epoch, modifying Eqs.~(\ref{phi}) and~(\ref{phinot}). 

For instance, within the scenario of active-sterile mixing \cite{Huang:2022wmz} where $\hat{y} \approx \sin^2 \theta g$, a potential $g \phi_{\odot} \sim 10^{-7}$ eV $(g / 10^{-22}) (10^{-18} \text{eV}/ m_{\phi})$ is able to maintain $\Delta N_{\text{eff}} \lesssim 10^{-2}$, increasing the potential will further reduce the contribution to $\Delta {N}_{\text{eff}}$. From the sensitivity to the ULSF parameter $\eta \sim$ 0.01 at ESS$\nu$SB, we obtain a similar potential $g \phi_{\odot} \sim 10^{-7}$ eV $(\hat{y} / 10^{-23}) (0.1/ \sin^2 \theta) (10^{-18} \text{eV}/ m_{\phi})$. Therefore, we do not expect a restriction from BBN in this scenario. Other constraints from BBN can be found in Refs.~\cite{Blinov:2019gcj,Dev:2022bae,Huang:2017egl,Venzor:2020ova}.

On the other hand, the CMB observations constrain the sum of the neutrino masses $\sum_{i} m_{i} \lesssim 0.23$ eV \cite{Planck:2015fie}; for modulations via $\Delta m_{31}^2$, according to~\cite{Krnjaic:2017zlz}, this implies $\eta(z=0) \lesssim 9 \times 10^{-3} ~(m_{\phi}/10^{-22}\text{eV})$, which is in good agreement with the ESS$\nu$SB sensitivity, $\eta(z=0)\lesssim 0.01$.

Within the CMB era, there could be an important case to consider for us. If the potential $ \hat{y }\phi$ dominates over the neutrino bare mass terms, the neutrino mass will significantly change. The relevant neutrino mass states $\nu_{1}, \nu_{3}$ will have the contribution~\cite{Berlin:2016woy}
\begin{equation}
\label{yphi}
    m_{1,3}(t) \simeq \hat{y}_{1,3} \langle \phi \rangle \simeq \frac{2 \hat{y}_{1,3}\sqrt{2\rho_{\phi}} }{ \pi m_{\phi}}.
\end{equation}
The CMB bound on the neutrino masses \footnote{Recently, fits using the CMB and LSS data sets give a constraint $\sum_{i}m_{i} \lesssim 0.40$ eV \cite{Lorenz:2021alz} that would imply a weaker restriction on $\hat{y}_{1,3}$.} implies a restriction for the coupling, 
$\hat{y}_{1,3}^{\text{CMB}}(z=0) \lesssim 6\times 10^{-21}~(m_{\phi}/10^{-22} \text{eV})$. As we have mentioned, ESS$\nu$SB will be sensitive to the phenomenological parameter $\eta$ [Eq. (\ref{eta})] at the percent level,
\begin{equation}
   \eta_{\Delta} \propto \hat{y}_{i}  \sqrt{ \rho_{\phi}} / m_{\phi}.
\end{equation}
This means that, for our usual choice of values $m_\phi$ and $\rho_{\phi}$, the sensitivity to the coupling $\hat{y}_{1,3}^{\text{CMB}}(z=0)$ would be at least one order of magnitude smaller than the ESS$\nu$SB sensitivity. 

The previous discussions assume that the ULSF accounts for all the DM in the Universe ($\rho_{\phi} = \rho_{\text{DM}, \odot}$). This may not be the case, and the ULSF could be only a fraction of this DM. 
The main phenomenological parameter, $\eta$, includes the DM density and would imply that $\hat{y}_{1,3} \propto \eta/\sqrt{\rho_{\phi}}$, if the $\phi$ accounts for less than 20$\%$ of the total DM density \cite{Kobayashi:2017jcf}; the scalar field amplitude $\phi_0$ decreases by a factor of two, thus the coupling $\hat{y}_{1,3} \gtrsim 2 \times 10^{-23}~(\eta/0.01) (m_{\phi} /10^{-22}~\text{eV})$. Therefore, in this scenario, the sensitivity to $\hat{y}_{1,3}$ will be weaker. 
On the other hand, for the local DM density $\rho_{\text{DM}, \odot} \simeq 0.3-0.6~ \text{GeV/cm}^3$; for $\rho_{\phi} =  \rho_{\text{DM}, \odot}$, the coupling can vary between the range $\hat{y}_{1,3}  \simeq \big[ 1/\sqrt{2} - 1 \big] \times10^{-23} ~(\eta/0.01)(m_{\phi}/10^{-22}\text{eV})$. 

Notice that, since $\hat{y}_{1,3} \propto \eta m_{\phi}$, a larger ULSF mass will reduce the sensitivity on the coupling $\hat{y}_{1,3}$ for fixed density $\rho_{\phi}$. However, most observational features of a ULSF as DM favor a mass range $m_{\phi} \sim (10^{-1}-10) \times 10^{-22}$ eV \cite{Urena-Lopez:2007cpl,Urena-Lopez:2019kud,Rodriguez-Montoya:2010ebl,Arbey:2001jj,Lesgourgues:2002hk,Arbey:2003sj,Matos:2000ng,Matos:2008ag,Arbey:2001qi,Matos:2000ss,Suarez:2011yf,Magana:2012ph,Suarez:2013iw,Hui:2016ltb}.

Finally, we might worry about the thermalization of the ULSF at distant past via $\nu \phi \rightarrow \nu \phi$ scatterings, which can lead to equilibrium among neutrinos and the ULSF. Nevertheless, the elastic scattering cross section for a scalar self-conjugate field, such as the ULSF with active neutrinos, via a spin one-half fermion $N$ ($M_{N} \gg m_{\phi}$) exchange is zero \cite{Boehm:2003hm}. \footnote{For the scenario discussed in Ref.~\cite{Dev:2022bae}, thermalization can be avoided.}

\section{Conclusions}
\label{conc}

The dark matter problem is one of the main puzzles in physics, and
different solutions have been proposed over the years. The existence
of dark matter has been a strong motivation to explore alternative
physical models and interactions to understand all its physical
consequences ranging from particle physics to cosmology. In
particular, the scalar field dark matter proposal is very successful
at the cosmological level and deserves further exploration of its
possible interactions with a Standard Model particle.

In this paper, we have discussed in detail the implications of a hypothetical interaction between neutrinos and an ULSF, as well as the case were this field is a DM candidate. We focused in
the case of the long-baseline neutrino experiment at the
ESS$\nu$SB. As already discussed, a broad set of proposals considers
either modifying of gravitational interactions or new particles at
different mass scales. We have focused on the case of an ultralight
scalar field and its hypothetical interaction with neutrinos. In this
scenario, a modification in the expected oscillation pattern in long
baseline neutrinos is expected. This modification could affect either
the-neutrino mixing angles or mass-squared differences. The ESS$\nu$SB
is sensitive to the ULSF via modulations on the
atmospheric mass squared difference $\Delta m_{31}^2$.

We found that sensitivities to the main parameter modeling the interaction between the ULSF and neutrinos are: $\eta
< 0.043 ~(0.039)$ at 3$\sigma$ and $\eta < 0.032 ~(0.028)$ at 90$\%$
C.L., from the $L=360$ km and ($L =540$ km) baselines,
respectively. Our bounds are comparable to other long-baseline
searches of this parameter. For instance, in
Ref.~\cite{Krnjaic:2017zlz}, a projected sensitivity of $\eta \sim
1.5 \%$ and $\eta \sim 7.5 \%$ via $\Delta m_{31}^2$ smearing are
expected at JUNO and DUNE experiments, respectively. Projected
1$\sigma$ sensitivities of $\eta \sim 0.035$ at
DUNE \cite{Dev:2020kgz} and $\eta \sim 0.01$ \cite{Losada:2021bxx} at
both DUNE and Hyper-Kamiokande were also reported in the
literature. Regarding reactor neutrino experiments, 1$\sigma$
sensitivities of $\eta \sim 0.08$ from Daya Bay and $\eta \sim 0.005$
from JUNO are reported \cite{Losada:2021bxx}. All these bounds are
comparable to our 1$\sigma$ sensitivities to the ULSF
parameter at the ESS$\nu$SB setup, namely $\eta \sim 0.03$ from the
$L= 360$ km baseline option and $\eta \sim 0.027$ from the $L= 540$ km
baseline. Therefore, the incorporation of the muon-neutrino
disappearance sample will not only benefit precision measurements at
ESS$\nu$SB \cite{ESSnuSB:2021azq} but also opens a window to search
for scalar-field modulations from the atmospheric mass-squared splitting $\Delta m_{31}^2$.

The ESS$\nu$SB experiment represents an opportunity to measure the
leptonic $CP$-violating phase accurately. Besides, it will allow
searching for different types of new physics at a competitive
level. The case of the ULSF candidate is an example of this potential.

\section*{Acknowledgments}
This work was partially supported by SNI-M\'exico and CONACyT research
Grant No.~A1-S-23238. Additionally the work of R.C. was partially supported
by COFAA-IPN, Estímulos al Desempeño de los Investigadores (EDI)-IPN and SIP-IPN Grants No.~20210704 and No. 20221329. We thank the anonymous referee for the illuminating comments, especially on the cosmological implications of the neutrino-ULSF interaction.



\begin{thebibliography}{93}
\expandafter\ifx\csname natexlab\endcsname\relax\def\natexlab#1{#1}\fi
\expandafter\ifx\csname bibnamefont\endcsname\relax
  \def\bibnamefont#1{#1}\fi
\expandafter\ifx\csname bibfnamefont\endcsname\relax
  \def\bibfnamefont#1{#1}\fi
\expandafter\ifx\csname citenamefont\endcsname\relax
  \def\citenamefont#1{#1}\fi
\expandafter\ifx\csname url\endcsname\relax
  \def\url#1{\texttt{#1}}\fi
\expandafter\ifx\csname urlprefix\endcsname\relax\def\urlprefix{URL }\fi
\providecommand{\bibinfo}[2]{#2}
\providecommand{\eprint}[2][]{\url{#2}}

\bibitem[{\citenamefont{Sahni and Wang}(2000)}]{Sahni:1999qe}
\bibinfo{author}{\bibfnamefont{V.}~\bibnamefont{Sahni}} \bibnamefont{and}
  \bibinfo{author}{\bibfnamefont{L.-M.} \bibnamefont{Wang}},
  \bibinfo{journal}{Phys. Rev. D} \textbf{\bibinfo{volume}{62}},
  \bibinfo{pages}{103517} (\bibinfo{year}{2000}), \eprint{astro-ph/9910097}.

\bibitem[{\citenamefont{Hu et~al.}(2000)\citenamefont{Hu, Barkana, and
  Gruzinov}}]{Hu:2000ke}
\bibinfo{author}{\bibfnamefont{W.}~\bibnamefont{Hu}},
  \bibinfo{author}{\bibfnamefont{R.}~\bibnamefont{Barkana}}, \bibnamefont{and}
  \bibinfo{author}{\bibfnamefont{A.}~\bibnamefont{Gruzinov}},
  \bibinfo{journal}{Phys. Rev. Lett.} \textbf{\bibinfo{volume}{85}},
  \bibinfo{pages}{1158} (\bibinfo{year}{2000}), \eprint{astro-ph/0003365}.

\bibitem[{\citenamefont{Matos et~al.}(2000)\citenamefont{Matos, Guzman, and
  Urena-Lopez}}]{Matos:1999et}
\bibinfo{author}{\bibfnamefont{T.}~\bibnamefont{Matos}},
  \bibinfo{author}{\bibfnamefont{F.~S.} \bibnamefont{Guzman}},
  \bibnamefont{and} \bibinfo{author}{\bibfnamefont{L.~A.}
  \bibnamefont{Urena-Lopez}}, \bibinfo{journal}{Class. Quant. Grav.}
  \textbf{\bibinfo{volume}{17}}, \bibinfo{pages}{1707} (\bibinfo{year}{2000}),
  \eprint{astro-ph/9908152}.

\bibitem[{\citenamefont{Matos and Urena-Lopez}(2000)}]{Matos:2000ng}
\bibinfo{author}{\bibfnamefont{T.}~\bibnamefont{Matos}} \bibnamefont{and}
  \bibinfo{author}{\bibfnamefont{L.~A.} \bibnamefont{Urena-Lopez}},
  \bibinfo{journal}{Class. Quant. Grav.} \textbf{\bibinfo{volume}{17}},
  \bibinfo{pages}{L75} (\bibinfo{year}{2000}), \eprint{astro-ph/0004332}.

\bibitem[{\citenamefont{Arbey et~al.}(2002)\citenamefont{Arbey, Lesgourgues,
  and Salati}}]{Arbey:2001jj}
\bibinfo{author}{\bibfnamefont{A.}~\bibnamefont{Arbey}},
  \bibinfo{author}{\bibfnamefont{J.}~\bibnamefont{Lesgourgues}},
  \bibnamefont{and} \bibinfo{author}{\bibfnamefont{P.}~\bibnamefont{Salati}},
  \bibinfo{journal}{Phys. Rev. D} \textbf{\bibinfo{volume}{65}},
  \bibinfo{pages}{083514} (\bibinfo{year}{2002}), \eprint{astro-ph/0112324}.

\bibitem[{\citenamefont{Arbey et~al.}(2001)\citenamefont{Arbey, Lesgourgues,
  and Salati}}]{Arbey:2001qi}
\bibinfo{author}{\bibfnamefont{A.}~\bibnamefont{Arbey}},
  \bibinfo{author}{\bibfnamefont{J.}~\bibnamefont{Lesgourgues}},
  \bibnamefont{and} \bibinfo{author}{\bibfnamefont{P.}~\bibnamefont{Salati}},
  \bibinfo{journal}{Phys. Rev. D} \textbf{\bibinfo{volume}{64}},
  \bibinfo{pages}{123528} (\bibinfo{year}{2001}), \eprint{astro-ph/0105564}.

\bibitem[{\citenamefont{Magana and Matos}(2012)}]{Magana:2012ph}
\bibinfo{author}{\bibfnamefont{J.}~\bibnamefont{Magana}} \bibnamefont{and}
  \bibinfo{author}{\bibfnamefont{T.}~\bibnamefont{Matos}}, \bibinfo{journal}{J.
  Phys. Conf. Ser.} \textbf{\bibinfo{volume}{378}}, \bibinfo{pages}{012012}
  (\bibinfo{year}{2012}), \eprint{1201.6107}.

\bibitem[{\citenamefont{Su\'arez et~al.}(2014)\citenamefont{Su\'arez, Robles,
  and Matos}}]{Suarez:2013iw}
\bibinfo{author}{\bibfnamefont{A.}~\bibnamefont{Su\'arez}},
  \bibinfo{author}{\bibfnamefont{V.~H.} \bibnamefont{Robles}},
  \bibnamefont{and} \bibinfo{author}{\bibfnamefont{T.}~\bibnamefont{Matos}},
  \bibinfo{journal}{Astrophys. Space Sci. Proc.} \textbf{\bibinfo{volume}{38}},
  \bibinfo{pages}{107} (\bibinfo{year}{2014}), \eprint{1302.0903}.

\bibitem[{\citenamefont{Marsh}(2016)}]{Marsh:2015xka}
\bibinfo{author}{\bibfnamefont{D.~J.~E.} \bibnamefont{Marsh}},
  \bibinfo{journal}{Phys. Rept.} \textbf{\bibinfo{volume}{643}},
  \bibinfo{pages}{1} (\bibinfo{year}{2016}), \eprint{1510.07633}.

\bibitem[{\citenamefont{Hui et~al.}(2017)\citenamefont{Hui, Ostriker, Tremaine,
  and Witten}}]{Hui:2016ltb}
\bibinfo{author}{\bibfnamefont{L.}~\bibnamefont{Hui}},
  \bibinfo{author}{\bibfnamefont{J.~P.} \bibnamefont{Ostriker}},
  \bibinfo{author}{\bibfnamefont{S.}~\bibnamefont{Tremaine}}, \bibnamefont{and}
  \bibinfo{author}{\bibfnamefont{E.}~\bibnamefont{Witten}},
  \bibinfo{journal}{Phys. Rev. D} \textbf{\bibinfo{volume}{95}},
  \bibinfo{pages}{043541} (\bibinfo{year}{2017}), \eprint{1610.08297}.

\bibitem[{\citenamefont{Lee}(2018)}]{Lee:2017qve}
\bibinfo{author}{\bibfnamefont{J.-W.} \bibnamefont{Lee}}, \bibinfo{journal}{EPJ
  Web Conf.} \textbf{\bibinfo{volume}{168}}, \bibinfo{pages}{06005}
  (\bibinfo{year}{2018}), \eprint{1704.05057}.

\bibitem[{\citenamefont{Ure\~na L\'opez}(2019)}]{Urena-Lopez:2019kud}
\bibinfo{author}{\bibfnamefont{L.~A.} \bibnamefont{Ure\~na L\'opez}},
  \bibinfo{journal}{Front. Astron. Space Sci.} \textbf{\bibinfo{volume}{6}},
  \bibinfo{pages}{47} (\bibinfo{year}{2019}).

\bibitem[{\citenamefont{Matos et~al.}(2009)\citenamefont{Matos,
  Vazquez-Gonzalez, and Magana}}]{Matos:2008ag}
\bibinfo{author}{\bibfnamefont{T.}~\bibnamefont{Matos}},
  \bibinfo{author}{\bibfnamefont{A.}~\bibnamefont{Vazquez-Gonzalez}},
  \bibnamefont{and} \bibinfo{author}{\bibfnamefont{J.}~\bibnamefont{Magana}},
  \bibinfo{journal}{Mon. Not. Roy. Astron. Soc.}
  \textbf{\bibinfo{volume}{393}}, \bibinfo{pages}{1359} (\bibinfo{year}{2009}),
  \eprint{0806.0683}.

\bibitem[{\citenamefont{Rodriguez-Montoya
  et~al.}(2010)\citenamefont{Rodriguez-Montoya, Magana, Matos, and
  Perez-Lorenzana}}]{Rodriguez-Montoya:2010ebl}
\bibinfo{author}{\bibfnamefont{I.}~\bibnamefont{Rodriguez-Montoya}},
  \bibinfo{author}{\bibfnamefont{J.}~\bibnamefont{Magana}},
  \bibinfo{author}{\bibfnamefont{T.}~\bibnamefont{Matos}}, \bibnamefont{and}
  \bibinfo{author}{\bibfnamefont{A.}~\bibnamefont{Perez-Lorenzana}},
  \bibinfo{journal}{Astrophys. J.} \textbf{\bibinfo{volume}{721}},
  \bibinfo{pages}{1509} (\bibinfo{year}{2010}), \eprint{0908.0054}.

\bibitem[{\citenamefont{Lesgourgues et~al.}(2002)\citenamefont{Lesgourgues,
  Arbey, and Salati}}]{Lesgourgues:2002hk}
\bibinfo{author}{\bibfnamefont{J.}~\bibnamefont{Lesgourgues}},
  \bibinfo{author}{\bibfnamefont{A.}~\bibnamefont{Arbey}}, \bibnamefont{and}
  \bibinfo{author}{\bibfnamefont{P.}~\bibnamefont{Salati}},
  \bibinfo{journal}{New Astron. Rev.} \textbf{\bibinfo{volume}{46}},
  \bibinfo{pages}{791} (\bibinfo{year}{2002}).

\bibitem[{\citenamefont{Arbey et~al.}(2003)\citenamefont{Arbey, Lesgourgues,
  and Salati}}]{Arbey:2003sj}
\bibinfo{author}{\bibfnamefont{A.}~\bibnamefont{Arbey}},
  \bibinfo{author}{\bibfnamefont{J.}~\bibnamefont{Lesgourgues}},
  \bibnamefont{and} \bibinfo{author}{\bibfnamefont{P.}~\bibnamefont{Salati}},
  \bibinfo{journal}{Phys. Rev. D} \textbf{\bibinfo{volume}{68}},
  \bibinfo{pages}{023511} (\bibinfo{year}{2003}), \eprint{astro-ph/0301533}.

\bibitem[{\citenamefont{Harko}(2011)}]{Harko:2011jy}
\bibinfo{author}{\bibfnamefont{T.}~\bibnamefont{Harko}}, \bibinfo{journal}{Mon.
  Not. Roy. Astron. Soc.} \textbf{\bibinfo{volume}{413}}, \bibinfo{pages}{3095}
  (\bibinfo{year}{2011}), \eprint{1101.3655}.

\bibitem[{\citenamefont{Robles and Matos}(2012)}]{Robles:2012uy}
\bibinfo{author}{\bibfnamefont{V.~H.} \bibnamefont{Robles}} \bibnamefont{and}
  \bibinfo{author}{\bibfnamefont{T.}~\bibnamefont{Matos}},
  \bibinfo{journal}{Mon. Not. Roy. Astron. Soc.}
  \textbf{\bibinfo{volume}{422}}, \bibinfo{pages}{282} (\bibinfo{year}{2012}),
  \eprint{1201.3032}.

\bibitem[{\citenamefont{Lee and Lim}(2010)}]{Lee:2008jp}
\bibinfo{author}{\bibfnamefont{J.-W.} \bibnamefont{Lee}} \bibnamefont{and}
  \bibinfo{author}{\bibfnamefont{S.}~\bibnamefont{Lim}},
  \bibinfo{journal}{JCAP} \textbf{\bibinfo{volume}{01}}, \bibinfo{pages}{007}
  (\bibinfo{year}{2010}), \eprint{0812.1342}.

\bibitem[{\citenamefont{Berlin}(2016)}]{Berlin:2016woy}
\bibinfo{author}{\bibfnamefont{A.}~\bibnamefont{Berlin}},
  \bibinfo{journal}{Phys. Rev. Lett.} \textbf{\bibinfo{volume}{117}},
  \bibinfo{pages}{231801} (\bibinfo{year}{2016}), \eprint{1608.01307}.

\bibitem[{\citenamefont{Brdar et~al.}(2018)\citenamefont{Brdar, Kopp, Liu,
  Prass, and Wang}}]{Brdar:2017kbt}
\bibinfo{author}{\bibfnamefont{V.}~\bibnamefont{Brdar}},
  \bibinfo{author}{\bibfnamefont{J.}~\bibnamefont{Kopp}},
  \bibinfo{author}{\bibfnamefont{J.}~\bibnamefont{Liu}},
  \bibinfo{author}{\bibfnamefont{P.}~\bibnamefont{Prass}}, \bibnamefont{and}
  \bibinfo{author}{\bibfnamefont{X.-P.} \bibnamefont{Wang}},
  \bibinfo{journal}{Phys. Rev. D} \textbf{\bibinfo{volume}{97}},
  \bibinfo{pages}{043001} (\bibinfo{year}{2018}), \eprint{1705.09455}.

\bibitem[{\citenamefont{Krnjaic et~al.}(2018)\citenamefont{Krnjaic, Machado,
  and Necib}}]{Krnjaic:2017zlz}
\bibinfo{author}{\bibfnamefont{G.}~\bibnamefont{Krnjaic}},
  \bibinfo{author}{\bibfnamefont{P.~A.~N.} \bibnamefont{Machado}},
  \bibnamefont{and} \bibinfo{author}{\bibfnamefont{L.}~\bibnamefont{Necib}},
  \bibinfo{journal}{Phys. Rev. D} \textbf{\bibinfo{volume}{97}},
  \bibinfo{pages}{075017} (\bibinfo{year}{2018}), \eprint{1705.06740}.

\bibitem[{\citenamefont{Liao et~al.}(2018)\citenamefont{Liao, Marfatia, and
  Whisnant}}]{Liao:2018byh}
\bibinfo{author}{\bibfnamefont{J.}~\bibnamefont{Liao}},
  \bibinfo{author}{\bibfnamefont{D.}~\bibnamefont{Marfatia}}, \bibnamefont{and}
  \bibinfo{author}{\bibfnamefont{K.}~\bibnamefont{Whisnant}},
  \bibinfo{journal}{JHEP} \textbf{\bibinfo{volume}{04}}, \bibinfo{pages}{136}
  (\bibinfo{year}{2018}), \eprint{1803.01773}.

\bibitem[{\citenamefont{Kim et~al.}(2020)\citenamefont{Kim, Machado, Park, and
  Shin}}]{Kim}
\bibinfo{author}{\bibfnamefont{D.}~\bibnamefont{Kim}},
  \bibinfo{author}{\bibfnamefont{P.~A.~N.} \bibnamefont{Machado}},
  \bibinfo{author}{\bibfnamefont{J.-C.} \bibnamefont{Park}}, \bibnamefont{and}
  \bibinfo{author}{\bibfnamefont{S.}~\bibnamefont{Shin}},
  \bibinfo{journal}{JHEP} \textbf{\bibinfo{volume}{07}}, \bibinfo{pages}{057}
  (\bibinfo{year}{2020}), \eprint{2003.07369}.

\bibitem[{\citenamefont{Dev et~al.}(2021)\citenamefont{Dev, Machado, and
  Mart\'\i{}nez-Mirav\'e}}]{Dev:2020kgz}
\bibinfo{author}{\bibfnamefont{A.}~\bibnamefont{Dev}},
  \bibinfo{author}{\bibfnamefont{P.~A.~N.} \bibnamefont{Machado}},
  \bibnamefont{and}
  \bibinfo{author}{\bibfnamefont{P.}~\bibnamefont{Mart\'\i{}nez-Mirav\'e}},
  \bibinfo{journal}{JHEP} \textbf{\bibinfo{volume}{01}}, \bibinfo{pages}{094}
  (\bibinfo{year}{2021}), \eprint{2007.03590}.

\bibitem[{\citenamefont{Losada et~al.}(2022{\natexlab{a}})\citenamefont{Losada,
  Nir, Perez, and Shpilman}}]{Losada:2021bxx}
\bibinfo{author}{\bibfnamefont{M.}~\bibnamefont{Losada}},
  \bibinfo{author}{\bibfnamefont{Y.}~\bibnamefont{Nir}},
  \bibinfo{author}{\bibfnamefont{G.}~\bibnamefont{Perez}}, \bibnamefont{and}
  \bibinfo{author}{\bibfnamefont{Y.}~\bibnamefont{Shpilman}},
  \bibinfo{journal}{JHEP} \textbf{\bibinfo{volume}{04}}, \bibinfo{pages}{030}
  (\bibinfo{year}{2022}{\natexlab{a}}), \eprint{2107.10865}.

\bibitem[{\citenamefont{Zhao}(2017)}]{Zhao:2017wmo}
\bibinfo{author}{\bibfnamefont{Y.}~\bibnamefont{Zhao}}, \bibinfo{journal}{Phys.
  Rev. D} \textbf{\bibinfo{volume}{95}}, \bibinfo{pages}{115002}
  (\bibinfo{year}{2017}), \eprint{1701.02735}.

\bibitem[{\citenamefont{Farzan}(2019)}]{Farzan}
\bibinfo{author}{\bibfnamefont{Y.}~\bibnamefont{Farzan}},
  \bibinfo{journal}{Phys. Lett. B} \textbf{\bibinfo{volume}{797}},
  \bibinfo{pages}{134911} (\bibinfo{year}{2019}), \eprint{1907.04271}.

\bibitem[{\citenamefont{Cline}(2020)}]{Cline:2019seo}
\bibinfo{author}{\bibfnamefont{J.~M.} \bibnamefont{Cline}},
  \bibinfo{journal}{Phys. Lett. B} \textbf{\bibinfo{volume}{802}},
  \bibinfo{pages}{135182} (\bibinfo{year}{2020}), \eprint{1908.02278}.

\bibitem[{\citenamefont{Dev et~al.}(2022)\citenamefont{Dev, Krnjaic, Machado,
  and Ramani}}]{Dev:2022bae}
\bibinfo{author}{\bibfnamefont{A.}~\bibnamefont{Dev}},
  \bibinfo{author}{\bibfnamefont{G.}~\bibnamefont{Krnjaic}},
  \bibinfo{author}{\bibfnamefont{P.}~\bibnamefont{Machado}}, \bibnamefont{and}
  \bibinfo{author}{\bibfnamefont{H.}~\bibnamefont{Ramani}}
  (\bibinfo{year}{2022}), \eprint{2205.06821}.

\bibitem[{\citenamefont{Huang et~al.}(2022)\citenamefont{Huang, Lindner,
  Mart\'\i{}nez-Mirav\'e, and Sen}}]{Huang:2022wmz}
\bibinfo{author}{\bibfnamefont{G.-y.} \bibnamefont{Huang}},
  \bibinfo{author}{\bibfnamefont{M.}~\bibnamefont{Lindner}},
  \bibinfo{author}{\bibfnamefont{P.}~\bibnamefont{Mart\'\i{}nez-Mirav\'e}},
  \bibnamefont{and} \bibinfo{author}{\bibfnamefont{M.}~\bibnamefont{Sen}},
  \bibinfo{journal}{Phys. Rev. D} \textbf{\bibinfo{volume}{106}},
  \bibinfo{pages}{033004} (\bibinfo{year}{2022}), \eprint{2205.08431}.

\bibitem[{\citenamefont{Losada et~al.}(2022{\natexlab{b}})\citenamefont{Losada,
  Nir, Perez, Savoray, and Shpilman}}]{Losada:2022uvr}
\bibinfo{author}{\bibfnamefont{M.}~\bibnamefont{Losada}},
  \bibinfo{author}{\bibfnamefont{Y.}~\bibnamefont{Nir}},
  \bibinfo{author}{\bibfnamefont{G.}~\bibnamefont{Perez}},
  \bibinfo{author}{\bibfnamefont{I.}~\bibnamefont{Savoray}}, \bibnamefont{and}
  \bibinfo{author}{\bibfnamefont{Y.}~\bibnamefont{Shpilman}}
  (\bibinfo{year}{2022}{\natexlab{b}}), \eprint{2205.09769}.


 \bibitem[{\citenamefont{Kobayashi et~al.}(2017)\citenamefont{Kobayashi, Murgia,
  De~Simone, Ir\v{s}i\v{c}, and Viel}}]{Kobayashi:2017jcf}
\bibinfo{author}{\bibfnamefont{T.}~\bibnamefont{Kobayashi}},
  \bibinfo{author}{\bibfnamefont{R.}~\bibnamefont{Murgia}},
  \bibinfo{author}{\bibfnamefont{A.}~\bibnamefont{De~Simone}},
  \bibinfo{author}{\bibfnamefont{V.}~\bibnamefont{Ir\v{s}i\v{c}}},
  \bibnamefont{and} \bibinfo{author}{\bibfnamefont{M.}~\bibnamefont{Viel}},
  \bibinfo{journal}{Phys. Rev. D} \textbf{\bibinfo{volume}{96}},
  \bibinfo{pages}{123514} (\bibinfo{year}{2017}), \eprint{1708.00015}.

  
\bibitem{Bar:2021kti}
N.~Bar, K.~Blum and C.~Sun,
Phys. Rev. D \textbf{105}, no.8, 8 (2022),
\eprint{2111.03070}. 

\bibitem[{\citenamefont{Baussan et~al.}(2014)}]{ESSnuSB:2013dql}
\bibinfo{author}{\bibfnamefont{E.}~\bibnamefont{Baussan}} \bibnamefont{et~al.}
  (\bibinfo{collaboration}{ESSnuSB}), \bibinfo{journal}{Nucl. Phys. B}
  \textbf{\bibinfo{volume}{885}}, \bibinfo{pages}{127} (\bibinfo{year}{2014}),
  \eprint{1309.7022}.

\bibitem[{\citenamefont{Alekou et~al.}(2021)}]{ESSnuSB:2021azq}
\bibinfo{author}{\bibfnamefont{A.}~\bibnamefont{Alekou}} \bibnamefont{et~al.}
  (\bibinfo{collaboration}{ESSnuSB}), \bibinfo{journal}{Eur. Phys. J. C}
  \textbf{\bibinfo{volume}{81}}, \bibinfo{pages}{1130} (\bibinfo{year}{2021}),
  \eprint{2107.07585}.

\bibitem[{\citenamefont{Alekou et~al.}(2022)}]{Alekou:2022emd}
\bibinfo{author}{\bibfnamefont{A.}~\bibnamefont{Alekou}} \bibnamefont{et~al.}
  (\bibinfo{year}{2022}), \eprint{2206.01208}.

\bibitem[{\citenamefont{Agostino et~al.}(2013)\citenamefont{Agostino,
  Buizza-Avanzini, Dracos, Duchesneau, Marafini, Mezzetto, Mosca, Patzak,
  Tonazzo, and Vassilopoulos}}]{MEMPHYS:2012bzz}
\bibinfo{author}{\bibfnamefont{L.}~\bibnamefont{Agostino}},
  \bibinfo{author}{\bibfnamefont{M.}~\bibnamefont{Buizza-Avanzini}},
  \bibinfo{author}{\bibfnamefont{M.}~\bibnamefont{Dracos}},
  \bibinfo{author}{\bibfnamefont{D.}~\bibnamefont{Duchesneau}},
  \bibinfo{author}{\bibfnamefont{M.}~\bibnamefont{Marafini}},
  \bibinfo{author}{\bibfnamefont{M.}~\bibnamefont{Mezzetto}},
  \bibinfo{author}{\bibfnamefont{L.}~\bibnamefont{Mosca}},
  \bibinfo{author}{\bibfnamefont{T.}~\bibnamefont{Patzak}},
  \bibinfo{author}{\bibfnamefont{A.}~\bibnamefont{Tonazzo}}, \bibnamefont{and}
  \bibinfo{author}{\bibfnamefont{N.}~\bibnamefont{Vassilopoulos}}
  (\bibinfo{collaboration}{MEMPHYS}), \bibinfo{journal}{JCAP}
  \textbf{\bibinfo{volume}{01}}, \bibinfo{pages}{024} (\bibinfo{year}{2013}),
  \eprint{1206.6665}.

\bibitem[{\citenamefont{Kumar~Agarwalla
  et~al.}(2019)\citenamefont{Kumar~Agarwalla, Chatterjee, and
  Palazzo}}]{KumarAgarwalla:2019blx}
\bibinfo{author}{\bibfnamefont{S.}~\bibnamefont{Kumar~Agarwalla}},
  \bibinfo{author}{\bibfnamefont{S.~S.} \bibnamefont{Chatterjee}},
  \bibnamefont{and} \bibinfo{author}{\bibfnamefont{A.}~\bibnamefont{Palazzo}},
  \bibinfo{journal}{JHEP} \textbf{\bibinfo{volume}{12}}, \bibinfo{pages}{174}
  (\bibinfo{year}{2019}), \eprint{1909.13746}.

\bibitem[{\citenamefont{Ghosh et~al.}(2020)\citenamefont{Ghosh, Ohlsson, and
  Rosauro-Alcaraz}}]{Ghosh:2019zvl}
\bibinfo{author}{\bibfnamefont{M.}~\bibnamefont{Ghosh}},
  \bibinfo{author}{\bibfnamefont{T.}~\bibnamefont{Ohlsson}}, \bibnamefont{and}
  \bibinfo{author}{\bibfnamefont{S.}~\bibnamefont{Rosauro-Alcaraz}},
  \bibinfo{journal}{JHEP} \textbf{\bibinfo{volume}{03}}, \bibinfo{pages}{026}
  (\bibinfo{year}{2020}), \eprint{1912.10010}.

\bibitem[{\citenamefont{Choubey et~al.}(2021)\citenamefont{Choubey, Ghosh,
  Kempe, and Ohlsson}}]{Choubey:2020dhw}
\bibinfo{author}{\bibfnamefont{S.}~\bibnamefont{Choubey}},
  \bibinfo{author}{\bibfnamefont{M.}~\bibnamefont{Ghosh}},
  \bibinfo{author}{\bibfnamefont{D.}~\bibnamefont{Kempe}}, \bibnamefont{and}
  \bibinfo{author}{\bibfnamefont{T.}~\bibnamefont{Ohlsson}},
  \bibinfo{journal}{JHEP} \textbf{\bibinfo{volume}{05}}, \bibinfo{pages}{133}
  (\bibinfo{year}{2021}), \eprint{2010.16334}.

\bibitem[{\citenamefont{Majhi et~al.}(2021)\citenamefont{Majhi, Singha,
  Deepthi, and Mohanta}}]{Majhi:2021api}
\bibinfo{author}{\bibfnamefont{R.}~\bibnamefont{Majhi}},
  \bibinfo{author}{\bibfnamefont{D.~K.} \bibnamefont{Singha}},
  \bibinfo{author}{\bibfnamefont{K.~N.} \bibnamefont{Deepthi}},
  \bibnamefont{and} \bibinfo{author}{\bibfnamefont{R.}~\bibnamefont{Mohanta}},
  \bibinfo{journal}{Phys. Rev. D} \textbf{\bibinfo{volume}{104}},
  \bibinfo{pages}{055002} (\bibinfo{year}{2021}), \eprint{2101.08202}.

\bibitem[{\citenamefont{Blennow
  et~al.}(2020{\natexlab{a}})\citenamefont{Blennow, Ghosh, Ohlsson, and
  Titov}}]{Blennow:2020snb}
\bibinfo{author}{\bibfnamefont{M.}~\bibnamefont{Blennow}},
  \bibinfo{author}{\bibfnamefont{M.}~\bibnamefont{Ghosh}},
  \bibinfo{author}{\bibfnamefont{T.}~\bibnamefont{Ohlsson}}, \bibnamefont{and}
  \bibinfo{author}{\bibfnamefont{A.}~\bibnamefont{Titov}},
  \bibinfo{journal}{JHEP} \textbf{\bibinfo{volume}{07}}, \bibinfo{pages}{014}
  (\bibinfo{year}{2020}{\natexlab{a}}), \eprint{2004.00017}.

\bibitem[{\citenamefont{Chatterjee et~al.}(2021)\citenamefont{Chatterjee,
  Miranda, T\'ortola, and Valle}}]{Chatterjee:2021xyu}
\bibinfo{author}{\bibfnamefont{S.~S.} \bibnamefont{Chatterjee}},
  \bibinfo{author}{\bibfnamefont{O.~G.} \bibnamefont{Miranda}},
  \bibinfo{author}{\bibfnamefont{M.}~\bibnamefont{T\'ortola}},
  \bibnamefont{and} \bibinfo{author}{\bibfnamefont{J.~W.~F.}
  \bibnamefont{Valle}} (\bibinfo{year}{2021}), \eprint{2111.08673}.

\bibitem[{\citenamefont{Ahn et~al.}(2022)\citenamefont{Ahn, Kang, Ramos, and
  Tanimoto}}]{Ahn:2022ufs}
\bibinfo{author}{\bibfnamefont{Y.~H.} \bibnamefont{Ahn}},
  \bibinfo{author}{\bibfnamefont{S.~K.} \bibnamefont{Kang}},
  \bibinfo{author}{\bibfnamefont{R.}~\bibnamefont{Ramos}}, \bibnamefont{and}
  \bibinfo{author}{\bibfnamefont{M.}~\bibnamefont{Tanimoto}}
  (\bibinfo{year}{2022}), \eprint{2205.02796}.

\bibitem[{\citenamefont{Huber et~al.}(2005)\citenamefont{Huber, Lindner, and
  Winter}}]{Huber:2004ka}
\bibinfo{author}{\bibfnamefont{P.}~\bibnamefont{Huber}},
  \bibinfo{author}{\bibfnamefont{M.}~\bibnamefont{Lindner}}, \bibnamefont{and}
  \bibinfo{author}{\bibfnamefont{W.}~\bibnamefont{Winter}},
  \bibinfo{journal}{Comput. Phys. Commun.} \textbf{\bibinfo{volume}{167}},
  \bibinfo{pages}{195} (\bibinfo{year}{2005}), \eprint{hep-ph/0407333}.

\bibitem[{\citenamefont{Huber et~al.}(2007)\citenamefont{Huber, Kopp, Lindner,
  Rolinec, and Winter}}]{Huber:2007ji}
\bibinfo{author}{\bibfnamefont{P.}~\bibnamefont{Huber}},
  \bibinfo{author}{\bibfnamefont{J.}~\bibnamefont{Kopp}},
  \bibinfo{author}{\bibfnamefont{M.}~\bibnamefont{Lindner}},
  \bibinfo{author}{\bibfnamefont{M.}~\bibnamefont{Rolinec}}, \bibnamefont{and}
  \bibinfo{author}{\bibfnamefont{W.}~\bibnamefont{Winter}},
  \bibinfo{journal}{Comput. Phys. Commun.} \textbf{\bibinfo{volume}{177}},
  \bibinfo{pages}{432} (\bibinfo{year}{2007}), \eprint{hep-ph/0701187}.

\bibitem[{\citenamefont{Carr and Kuhnel}(2020)}]{Carr:2020xqk}
\bibinfo{author}{\bibfnamefont{B.}~\bibnamefont{Carr}} \bibnamefont{and}
  \bibinfo{author}{\bibfnamefont{F.}~\bibnamefont{Kuhnel}},
  \bibinfo{journal}{Ann. Rev. Nucl. Part. Sci.} \textbf{\bibinfo{volume}{70}},
  \bibinfo{pages}{355} (\bibinfo{year}{2020}), \eprint{2006.02838}.

\bibitem[{\citenamefont{Carr et~al.}(2021)\citenamefont{Carr, Kuhnel, and
  Visinelli}}]{Carr:2020mqm}
\bibinfo{author}{\bibfnamefont{B.}~\bibnamefont{Carr}},
  \bibinfo{author}{\bibfnamefont{F.}~\bibnamefont{Kuhnel}}, \bibnamefont{and}
  \bibinfo{author}{\bibfnamefont{L.}~\bibnamefont{Visinelli}},
  \bibinfo{journal}{Mon. Not. Roy. Astron. Soc.}
  \textbf{\bibinfo{volume}{506}}, \bibinfo{pages}{3648} (\bibinfo{year}{2021}),
  \eprint{2011.01930}.

\bibitem[{\citenamefont{Arbey and Mahmoudi}(2021)}]{Arbey:2021gdg}
\bibinfo{author}{\bibfnamefont{A.}~\bibnamefont{Arbey}} \bibnamefont{and}
  \bibinfo{author}{\bibfnamefont{F.}~\bibnamefont{Mahmoudi}},
  \bibinfo{journal}{Prog. Part. Nucl. Phys.} \textbf{\bibinfo{volume}{119}},
  \bibinfo{pages}{103865} (\bibinfo{year}{2021}), \eprint{2104.11488}.

\bibitem[{\citenamefont{Di~Luzio et~al.}(2020)\citenamefont{Di~Luzio,
  Giannotti, Nardi, and Visinelli}}]{DiLuzio:2020wdo}
\bibinfo{author}{\bibfnamefont{L.}~\bibnamefont{Di~Luzio}},
  \bibinfo{author}{\bibfnamefont{M.}~\bibnamefont{Giannotti}},
  \bibinfo{author}{\bibfnamefont{E.}~\bibnamefont{Nardi}}, \bibnamefont{and}
  \bibinfo{author}{\bibfnamefont{L.}~\bibnamefont{Visinelli}},
  \bibinfo{journal}{Phys. Rept.} \textbf{\bibinfo{volume}{870}},
  \bibinfo{pages}{1} (\bibinfo{year}{2020}), \eprint{2003.01100}.

\bibitem[{\citenamefont{Arun et~al.}(2017)\citenamefont{Arun, Gudennavar, and
  Sivaram}}]{Arun:2017uaw}
\bibinfo{author}{\bibfnamefont{K.}~\bibnamefont{Arun}},
  \bibinfo{author}{\bibfnamefont{S.~B.} \bibnamefont{Gudennavar}},
  \bibnamefont{and} \bibinfo{author}{\bibfnamefont{C.}~\bibnamefont{Sivaram}},
  \bibinfo{journal}{Adv. Space Res.} \textbf{\bibinfo{volume}{60}},
  \bibinfo{pages}{166} (\bibinfo{year}{2017}), \eprint{1704.06155}.

\bibitem[{\citenamefont{Drewes}(2013)}]{Drewes:2013gca}
\bibinfo{author}{\bibfnamefont{M.}~\bibnamefont{Drewes}},
  \bibinfo{journal}{Int. J. Mod. Phys. E} \textbf{\bibinfo{volume}{22}},
  \bibinfo{pages}{1330019} (\bibinfo{year}{2013}), \eprint{1303.6912}.

\bibitem[{\citenamefont{Ng et~al.}(2019)\citenamefont{Ng, Roach, Perez, Beacom,
  Horiuchi, Krivonos, and Wik}}]{Ng:2019gch}
\bibinfo{author}{\bibfnamefont{K.~C.~Y.} \bibnamefont{Ng}},
  \bibinfo{author}{\bibfnamefont{B.~M.} \bibnamefont{Roach}},
  \bibinfo{author}{\bibfnamefont{K.}~\bibnamefont{Perez}},
  \bibinfo{author}{\bibfnamefont{J.~F.} \bibnamefont{Beacom}},
  \bibinfo{author}{\bibfnamefont{S.}~\bibnamefont{Horiuchi}},
  \bibinfo{author}{\bibfnamefont{R.}~\bibnamefont{Krivonos}}, \bibnamefont{and}
  \bibinfo{author}{\bibfnamefont{D.~R.} \bibnamefont{Wik}},
  \bibinfo{journal}{Phys. Rev. D} \textbf{\bibinfo{volume}{99}},
  \bibinfo{pages}{083005} (\bibinfo{year}{2019}), \eprint{1901.01262}.

\bibitem[{\citenamefont{Oks}(2021)}]{Oks:2021hef}
\bibinfo{author}{\bibfnamefont{E.}~\bibnamefont{Oks}}, \bibinfo{journal}{New
  Astron. Rev.} \textbf{\bibinfo{volume}{93}}, \bibinfo{pages}{101632}
  (\bibinfo{year}{2021}), \eprint{2111.00363}.

\bibitem[{\citenamefont{Milgrom}(1983)}]{Milgrom:1983ca}
\bibinfo{author}{\bibfnamefont{M.}~\bibnamefont{Milgrom}},
  \bibinfo{journal}{Astrophys. J.} \textbf{\bibinfo{volume}{270}},
  \bibinfo{pages}{365} (\bibinfo{year}{1983}).

\bibitem[{\citenamefont{Baker et~al.}(2021)}]{Baker:2019gxo}
\bibinfo{author}{\bibfnamefont{T.}~\bibnamefont{Baker}} \bibnamefont{et~al.},
  \bibinfo{journal}{Rev. Mod. Phys.} \textbf{\bibinfo{volume}{93}},
  \bibinfo{pages}{015003} (\bibinfo{year}{2021}), \eprint{1908.03430}.

\bibitem[{\citenamefont{Randall and
  Sundrum}(1999{\natexlab{a}})}]{Randall:1999ee}
\bibinfo{author}{\bibfnamefont{L.}~\bibnamefont{Randall}} \bibnamefont{and}
  \bibinfo{author}{\bibfnamefont{R.}~\bibnamefont{Sundrum}},
  \bibinfo{journal}{Phys. Rev. Lett.} \textbf{\bibinfo{volume}{83}},
  \bibinfo{pages}{3370} (\bibinfo{year}{1999}{\natexlab{a}}),
  \eprint{hep-ph/9905221}.

\bibitem[{\citenamefont{Randall and
  Sundrum}(1999{\natexlab{b}})}]{Randall:1999vf}
\bibinfo{author}{\bibfnamefont{L.}~\bibnamefont{Randall}} \bibnamefont{and}
  \bibinfo{author}{\bibfnamefont{R.}~\bibnamefont{Sundrum}},
  \bibinfo{journal}{Phys. Rev. Lett.} \textbf{\bibinfo{volume}{83}},
  \bibinfo{pages}{4690} (\bibinfo{year}{1999}{\natexlab{b}}),
  \eprint{hep-th/9906064}.

\bibitem[{\citenamefont{Clowe et~al.}(2006)\citenamefont{Clowe, Bradac,
  Gonzalez, Markevitch, Randall, Jones, and Zaritsky}}]{Clowe:2006eq}
\bibinfo{author}{\bibfnamefont{D.}~\bibnamefont{Clowe}},
  \bibinfo{author}{\bibfnamefont{M.}~\bibnamefont{Bradac}},
  \bibinfo{author}{\bibfnamefont{A.~H.} \bibnamefont{Gonzalez}},
  \bibinfo{author}{\bibfnamefont{M.}~\bibnamefont{Markevitch}},
  \bibinfo{author}{\bibfnamefont{S.~W.} \bibnamefont{Randall}},
  \bibinfo{author}{\bibfnamefont{C.}~\bibnamefont{Jones}}, \bibnamefont{and}
  \bibinfo{author}{\bibfnamefont{D.}~\bibnamefont{Zaritsky}},
  \bibinfo{journal}{Astrophys. J. Lett.} \textbf{\bibinfo{volume}{648}},
  \bibinfo{pages}{L109} (\bibinfo{year}{2006}), \eprint{astro-ph/0608407}.

\bibitem[{\citenamefont{Klypin et~al.}(1999)\citenamefont{Klypin, Kravtsov,
  Valenzuela, and Prada}}]{Klypin:1999uc}
\bibinfo{author}{\bibfnamefont{A.~A.} \bibnamefont{Klypin}},
  \bibinfo{author}{\bibfnamefont{A.~V.} \bibnamefont{Kravtsov}},
  \bibinfo{author}{\bibfnamefont{O.}~\bibnamefont{Valenzuela}},
  \bibnamefont{and} \bibinfo{author}{\bibfnamefont{F.}~\bibnamefont{Prada}},
  \bibinfo{journal}{Astrophys. J.} \textbf{\bibinfo{volume}{522}},
  \bibinfo{pages}{82} (\bibinfo{year}{1999}), \eprint{astro-ph/9901240}.

\bibitem[{\citenamefont{Moore et~al.}(1999)\citenamefont{Moore, Ghigna,
  Governato, Lake, Quinn, Stadel, and Tozzi}}]{Moore:1999nt}
\bibinfo{author}{\bibfnamefont{B.}~\bibnamefont{Moore}},
  \bibinfo{author}{\bibfnamefont{S.}~\bibnamefont{Ghigna}},
  \bibinfo{author}{\bibfnamefont{F.}~\bibnamefont{Governato}},
  \bibinfo{author}{\bibfnamefont{G.}~\bibnamefont{Lake}},
  \bibinfo{author}{\bibfnamefont{T.~R.} \bibnamefont{Quinn}},
  \bibinfo{author}{\bibfnamefont{J.}~\bibnamefont{Stadel}}, \bibnamefont{and}
  \bibinfo{author}{\bibfnamefont{P.}~\bibnamefont{Tozzi}},
  \bibinfo{journal}{Astrophys. J. Lett.} \textbf{\bibinfo{volume}{524}},
  \bibinfo{pages}{L19} (\bibinfo{year}{1999}), \eprint{astro-ph/9907411}.

\bibitem[{\citenamefont{Peebles and Nusser}(2010)}]{Peebles:2010di}
\bibinfo{author}{\bibfnamefont{P.~J.~E.} \bibnamefont{Peebles}}
  \bibnamefont{and} \bibinfo{author}{\bibfnamefont{A.}~\bibnamefont{Nusser}},
  \bibinfo{journal}{Nature} \textbf{\bibinfo{volume}{465}},
  \bibinfo{pages}{565} (\bibinfo{year}{2010}), \eprint{1001.1484}.

\bibitem[{\citenamefont{Lundgren et~al.}(2010)\citenamefont{Lundgren,
  Bondarescu, Bondarescu, and Balakrishna}}]{Lundgren:2010sp}
\bibinfo{author}{\bibfnamefont{A.~P.} \bibnamefont{Lundgren}},
  \bibinfo{author}{\bibfnamefont{M.}~\bibnamefont{Bondarescu}},
  \bibinfo{author}{\bibfnamefont{R.}~\bibnamefont{Bondarescu}},
  \bibnamefont{and}
  \bibinfo{author}{\bibfnamefont{J.}~\bibnamefont{Balakrishna}},
  \bibinfo{journal}{Astrophys. J. Lett.} \textbf{\bibinfo{volume}{715}},
  \bibinfo{pages}{L35} (\bibinfo{year}{2010}), \eprint{1001.0051}.

\bibitem[{\citenamefont{Matos and Urena-Lopez}(2001)}]{Matos:2000ss}
\bibinfo{author}{\bibfnamefont{T.}~\bibnamefont{Matos}} \bibnamefont{and}
  \bibinfo{author}{\bibfnamefont{L.~A.} \bibnamefont{Urena-Lopez}},
  \bibinfo{journal}{Phys. Rev. D} \textbf{\bibinfo{volume}{63}},
  \bibinfo{pages}{063506} (\bibinfo{year}{2001}), \eprint{astro-ph/0006024}.

\bibitem[{\citenamefont{Suarez and Matos}(2011)}]{Suarez:2011yf}
\bibinfo{author}{\bibfnamefont{A.}~\bibnamefont{Suarez}} \bibnamefont{and}
  \bibinfo{author}{\bibfnamefont{T.}~\bibnamefont{Matos}},
  \bibinfo{journal}{Mon. Not. Roy. Astron. Soc.}
  \textbf{\bibinfo{volume}{416}}, \bibinfo{pages}{87} (\bibinfo{year}{2011}),
  \eprint{1101.4039}.

\bibitem[{\citenamefont{Arvanitaki et~al.}(2010)\citenamefont{Arvanitaki,
  Dimopoulos, Dubovsky, Kaloper, and March-Russell}}]{Arvanitaki:2009fg}
\bibinfo{author}{\bibfnamefont{A.}~\bibnamefont{Arvanitaki}},
  \bibinfo{author}{\bibfnamefont{S.}~\bibnamefont{Dimopoulos}},
  \bibinfo{author}{\bibfnamefont{S.}~\bibnamefont{Dubovsky}},
  \bibinfo{author}{\bibfnamefont{N.}~\bibnamefont{Kaloper}}, \bibnamefont{and}
  \bibinfo{author}{\bibfnamefont{J.}~\bibnamefont{March-Russell}},
  \bibinfo{journal}{Phys. Rev. D} \textbf{\bibinfo{volume}{81}},
  \bibinfo{pages}{123530} (\bibinfo{year}{2010}), \eprint{0905.4720}.

\bibitem[{\citenamefont{Cicoli et~al.}(2022)\citenamefont{Cicoli, Guidetti,
  Righi, and Westphal}}]{Cicoli:2021gss}
\bibinfo{author}{\bibfnamefont{M.}~\bibnamefont{Cicoli}},
  \bibinfo{author}{\bibfnamefont{V.}~\bibnamefont{Guidetti}},
  \bibinfo{author}{\bibfnamefont{N.}~\bibnamefont{Righi}}, \bibnamefont{and}
  \bibinfo{author}{\bibfnamefont{A.}~\bibnamefont{Westphal}},
  \bibinfo{journal}{JHEP} \textbf{\bibinfo{volume}{05}}, \bibinfo{pages}{107}
  (\bibinfo{year}{2022}), \eprint{2110.02964}.

\bibitem[{\citenamefont{Bernal et~al.}(2018)\citenamefont{Bernal,
  Fern\'andez-Hern\'andez, Matos, and Rodr\'\i{}guez-Meza}}]{Bernal:2017oih}
\bibinfo{author}{\bibfnamefont{T.}~\bibnamefont{Bernal}},
  \bibinfo{author}{\bibfnamefont{L.~M.} \bibnamefont{Fern\'andez-Hern\'andez}},
  \bibinfo{author}{\bibfnamefont{T.}~\bibnamefont{Matos}}, \bibnamefont{and}
  \bibinfo{author}{\bibfnamefont{M.~A.} \bibnamefont{Rodr\'\i{}guez-Meza}},
  \bibinfo{journal}{Mon. Not. Roy. Astron. Soc.}
  \textbf{\bibinfo{volume}{475}}, \bibinfo{pages}{1447} (\bibinfo{year}{2018}),
  \eprint{1701.00912}.

\bibitem[{\citenamefont{Ure\~na L\'opez et~al.}(2017)\citenamefont{Ure\~na
  L\'opez, Robles, and Matos}}]{Urena-Lopez:2017tob}
\bibinfo{author}{\bibfnamefont{L.~A.} \bibnamefont{Ure\~na L\'opez}},
  \bibinfo{author}{\bibfnamefont{V.~H.} \bibnamefont{Robles}},
  \bibnamefont{and} \bibinfo{author}{\bibfnamefont{T.}~\bibnamefont{Matos}},
  \bibinfo{journal}{Phys. Rev. D} \textbf{\bibinfo{volume}{96}},
  \bibinfo{pages}{043005} (\bibinfo{year}{2017}), \eprint{1702.05103}.

\bibitem[{\citenamefont{Hlozek et~al.}(2015)\citenamefont{Hlozek, Grin, Marsh,
  and Ferreira}}]{Hlozek:2014lca}
\bibinfo{author}{\bibfnamefont{R.}~\bibnamefont{Hlozek}},
  \bibinfo{author}{\bibfnamefont{D.}~\bibnamefont{Grin}},
  \bibinfo{author}{\bibfnamefont{D.~J.~E.} \bibnamefont{Marsh}},
  \bibnamefont{and} \bibinfo{author}{\bibfnamefont{P.~G.}
  \bibnamefont{Ferreira}}, \bibinfo{journal}{Phys. Rev. D}
  \textbf{\bibinfo{volume}{91}}, \bibinfo{pages}{103512}
  (\bibinfo{year}{2015}), \eprint{1410.2896}.



\bibitem{Bar:2018acw}
N.~Bar, D.~Blas, K.~Blum and S.~Sibiryakov,
Phys. Rev. D \textbf{98}, no.8, 083027 (2018),
\eprint{1805.00122}.




\bibitem[{\citenamefont{Arvanitaki et~al.}(2015)\citenamefont{Arvanitaki,
  Huang, and Van~Tilburg}}]{Arvanitaki:2014faa}
\bibinfo{author}{\bibfnamefont{A.}~\bibnamefont{Arvanitaki}},
  \bibinfo{author}{\bibfnamefont{J.}~\bibnamefont{Huang}}, \bibnamefont{and}
  \bibinfo{author}{\bibfnamefont{K.}~\bibnamefont{Van~Tilburg}},
  \bibinfo{journal}{Phys. Rev. D} \textbf{\bibinfo{volume}{91}},
  \bibinfo{pages}{015015} (\bibinfo{year}{2015}), \eprint{1405.2925}.

\bibitem[{\citenamefont{Urena-Lopez}(2007)}]{Urena-Lopez:2007cpl}
\bibinfo{author}{\bibfnamefont{L.~A.} \bibnamefont{Urena-Lopez}}, in
  \emph{\bibinfo{booktitle}{{4th Mexican School of Astrophysics}}}
  (\bibinfo{year}{2007}), pp. \bibinfo{pages}{295--302}.

\bibitem[{\citenamefont{de~Salas et~al.}(2019)\citenamefont{de~Salas, Malhan,
  Freese, Hattori, and Valluri}}]{deSalas:2019pee}
\bibinfo{author}{\bibfnamefont{P.~F.} \bibnamefont{de~Salas}},
  \bibinfo{author}{\bibfnamefont{K.}~\bibnamefont{Malhan}},
  \bibinfo{author}{\bibfnamefont{K.}~\bibnamefont{Freese}},
  \bibinfo{author}{\bibfnamefont{K.}~\bibnamefont{Hattori}}, \bibnamefont{and}
  \bibinfo{author}{\bibfnamefont{M.}~\bibnamefont{Valluri}},
  \bibinfo{journal}{JCAP} \textbf{\bibinfo{volume}{10}}, \bibinfo{pages}{037}
  (\bibinfo{year}{2019}), \eprint{1906.06133}.

\bibitem[{\citenamefont{de~Salas and Widmark}(2021)}]{deSalas:2020hbh}
\bibinfo{author}{\bibfnamefont{P.~F.} \bibnamefont{de~Salas}} \bibnamefont{and}
  \bibinfo{author}{\bibfnamefont{A.}~\bibnamefont{Widmark}},
  \bibinfo{journal}{Rept. Prog. Phys.} \textbf{\bibinfo{volume}{84}},
  \bibinfo{pages}{104901} (\bibinfo{year}{2021}), \eprint{2012.11477}.

\bibitem[{\citenamefont{Sivertsson et~al.}(2022)\citenamefont{Sivertsson, Read,
  Silverwood, de~Salas, Malhan, Widmark, Laporte, Garbari, and
  Freese}}]{Sivertsson:2022riu}
\bibinfo{author}{\bibfnamefont{S.}~\bibnamefont{Sivertsson}},
  \bibinfo{author}{\bibfnamefont{J.~I.} \bibnamefont{Read}},
  \bibinfo{author}{\bibfnamefont{H.}~\bibnamefont{Silverwood}},
  \bibinfo{author}{\bibfnamefont{P.~F.} \bibnamefont{de~Salas}},
  \bibinfo{author}{\bibfnamefont{K.}~\bibnamefont{Malhan}},
  \bibinfo{author}{\bibfnamefont{A.}~\bibnamefont{Widmark}},
  \bibinfo{author}{\bibfnamefont{C.~F.~P.} \bibnamefont{Laporte}},
  \bibinfo{author}{\bibfnamefont{S.}~\bibnamefont{Garbari}}, \bibnamefont{and}
  \bibinfo{author}{\bibfnamefont{K.}~\bibnamefont{Freese}},
  \bibinfo{journal}{Mon. Not. Roy. Astron. Soc.}
  \textbf{\bibinfo{volume}{511}}, \bibinfo{pages}{1977} (\bibinfo{year}{2022}),
  \eprint{2201.01822}.

\bibitem[{\citenamefont{Campagne et~al.}(2007)\citenamefont{Campagne, Maltoni,
  Mezzetto, and Schwetz}}]{Campagne:2006yx}
\bibinfo{author}{\bibfnamefont{J.-E.} \bibnamefont{Campagne}},
  \bibinfo{author}{\bibfnamefont{M.}~\bibnamefont{Maltoni}},
  \bibinfo{author}{\bibfnamefont{M.}~\bibnamefont{Mezzetto}}, \bibnamefont{and}
  \bibinfo{author}{\bibfnamefont{T.}~\bibnamefont{Schwetz}},
  \bibinfo{journal}{JHEP} \textbf{\bibinfo{volume}{04}}, \bibinfo{pages}{003}
  (\bibinfo{year}{2007}), \eprint{hep-ph/0603172}.

\bibitem[{\citenamefont{Blennow
  et~al.}(2020{\natexlab{b}})\citenamefont{Blennow, Fernandez-Martinez, Ota,
  and Rosauro-Alcaraz}}]{Blennow:2019bvl}
\bibinfo{author}{\bibfnamefont{M.}~\bibnamefont{Blennow}},
  \bibinfo{author}{\bibfnamefont{E.}~\bibnamefont{Fernandez-Martinez}},
  \bibinfo{author}{\bibfnamefont{T.}~\bibnamefont{Ota}}, \bibnamefont{and}
  \bibinfo{author}{\bibfnamefont{S.}~\bibnamefont{Rosauro-Alcaraz}},
  \bibinfo{journal}{Eur. Phys. J. C} \textbf{\bibinfo{volume}{80}},
  \bibinfo{pages}{190} (\bibinfo{year}{2020}{\natexlab{b}}),
  \eprint{1912.04309}.

\bibitem[{\citenamefont{Nunokawa et~al.}(2008)\citenamefont{Nunokawa, Parke,
  and Valle}}]{Nunokawa:2007qh}
\bibinfo{author}{\bibfnamefont{H.}~\bibnamefont{Nunokawa}},
  \bibinfo{author}{\bibfnamefont{S.~J.} \bibnamefont{Parke}}, \bibnamefont{and}
  \bibinfo{author}{\bibfnamefont{J.~W.~F.} \bibnamefont{Valle}},
  \bibinfo{journal}{Prog. Part. Nucl. Phys.} \textbf{\bibinfo{volume}{60}},
  \bibinfo{pages}{338} (\bibinfo{year}{2008}), \eprint{0710.0554}.

\bibitem[{\citenamefont{de~Salas et~al.}(2021)\citenamefont{de~Salas, Forero,
  Gariazzo, Mart\'\i{}nez-Mirav\'e, Mena, Ternes, T\'ortola, and
  Valle}}]{deSalas:2020pgw}
\bibinfo{author}{\bibfnamefont{P.~F.} \bibnamefont{de~Salas}},
  \bibinfo{author}{\bibfnamefont{D.~V.} \bibnamefont{Forero}},
  \bibinfo{author}{\bibfnamefont{S.}~\bibnamefont{Gariazzo}},
  \bibinfo{author}{\bibfnamefont{P.}~\bibnamefont{Mart\'\i{}nez-Mirav\'e}},
  \bibinfo{author}{\bibfnamefont{O.}~\bibnamefont{Mena}},
  \bibinfo{author}{\bibfnamefont{C.~A.} \bibnamefont{Ternes}},
  \bibinfo{author}{\bibfnamefont{M.}~\bibnamefont{T\'ortola}},
  \bibnamefont{and} \bibinfo{author}{\bibfnamefont{J.~W.~F.}
  \bibnamefont{Valle}}, \bibinfo{journal}{JHEP} \textbf{\bibinfo{volume}{02}},
  \bibinfo{pages}{071} (\bibinfo{year}{2021}), \eprint{2006.11237}.

\bibitem[{\citenamefont{Esteban et~al.}(2020)\citenamefont{Esteban,
  Gonzalez-Garcia, Maltoni, Schwetz, and Zhou}}]{Esteban:2020cvm}
\bibinfo{author}{\bibfnamefont{I.}~\bibnamefont{Esteban}},
  \bibinfo{author}{\bibfnamefont{M.~C.} \bibnamefont{Gonzalez-Garcia}},
  \bibinfo{author}{\bibfnamefont{M.}~\bibnamefont{Maltoni}},
  \bibinfo{author}{\bibfnamefont{T.}~\bibnamefont{Schwetz}}, \bibnamefont{and}
  \bibinfo{author}{\bibfnamefont{A.}~\bibnamefont{Zhou}},
  \bibinfo{journal}{JHEP} \textbf{\bibinfo{volume}{09}}, \bibinfo{pages}{178}
  (\bibinfo{year}{2020}), \eprint{2007.14792}.

\bibitem[{\citenamefont{Agarwalla et~al.}(2014)\citenamefont{Agarwalla,
  Choubey, and Prakash}}]{Agarwalla:2014tpa}
\bibinfo{author}{\bibfnamefont{S.~K.} \bibnamefont{Agarwalla}},
  \bibinfo{author}{\bibfnamefont{S.}~\bibnamefont{Choubey}}, \bibnamefont{and}
  \bibinfo{author}{\bibfnamefont{S.}~\bibnamefont{Prakash}},
  \bibinfo{journal}{JHEP} \textbf{\bibinfo{volume}{12}}, \bibinfo{pages}{020}
  (\bibinfo{year}{2014}), \eprint{1406.2219}.

\bibitem[{\citenamefont{Chakraborty et~al.}(2019)\citenamefont{Chakraborty,
  Goswami, Gupta, and Thakore}}]{Chakraborty:2019jlv}
\bibinfo{author}{\bibfnamefont{K.}~\bibnamefont{Chakraborty}},
  \bibinfo{author}{\bibfnamefont{S.}~\bibnamefont{Goswami}},
  \bibinfo{author}{\bibfnamefont{C.}~\bibnamefont{Gupta}}, \bibnamefont{and}
  \bibinfo{author}{\bibfnamefont{T.}~\bibnamefont{Thakore}},
  \bibinfo{journal}{JHEP} \textbf{\bibinfo{volume}{05}}, \bibinfo{pages}{137}
  (\bibinfo{year}{2019}), \eprint{1902.02963}.

\bibitem[{\citenamefont{Choubey and Roy}(2006)}]{Choubey:2005zy}
\bibinfo{author}{\bibfnamefont{S.}~\bibnamefont{Choubey}} \bibnamefont{and}
  \bibinfo{author}{\bibfnamefont{P.}~\bibnamefont{Roy}},
  \bibinfo{journal}{Phys. Rev. D} \textbf{\bibinfo{volume}{73}},
  \bibinfo{pages}{013006} (\bibinfo{year}{2006}), \eprint{hep-ph/0509197}.

\bibitem[{\citenamefont{Huber et~al.}(2002)\citenamefont{Huber, Lindner, and
  Winter}}]{Huber:2002mx}
\bibinfo{author}{\bibfnamefont{P.}~\bibnamefont{Huber}},
  \bibinfo{author}{\bibfnamefont{M.}~\bibnamefont{Lindner}}, \bibnamefont{and}
  \bibinfo{author}{\bibfnamefont{W.}~\bibnamefont{Winter}},
  \bibinfo{journal}{Nucl. Phys. B} \textbf{\bibinfo{volume}{645}},
  \bibinfo{pages}{3} (\bibinfo{year}{2002}), \eprint{hep-ph/0204352}.


\bibitem[{\citenamefont{Ade et~al.}(2016)}]{Planck:2015fie}
\bibinfo{author}{\bibfnamefont{P.~A.~R.} \bibnamefont{Ade}}
  \bibnamefont{et~al.} (\bibinfo{collaboration}{Planck}),
  \bibinfo{journal}{Astron. Astrophys.} \textbf{\bibinfo{volume}{594}},
  \bibinfo{pages}{A13} (\bibinfo{year}{2016}), \eprint{1502.01589}.


\bibitem{Capozzi:2018bps}
F.~Capozzi, I.~M.~Shoemaker and L.~Vecchi,
JCAP \textbf{07}, 004 (2018),
\eprint{1804.05117}.


\bibitem[{\citenamefont{Blinov et~al.}(2019)\citenamefont{Blinov, Kelly,
  Krnjaic, and McDermott}}]{Blinov:2019gcj}
\bibinfo{author}{\bibfnamefont{N.}~\bibnamefont{Blinov}},
  \bibinfo{author}{\bibfnamefont{K.~J.} \bibnamefont{Kelly}},
  \bibinfo{author}{\bibfnamefont{G.~Z.} \bibnamefont{Krnjaic}},
  \bibnamefont{and} \bibinfo{author}{\bibfnamefont{S.~D.}
  \bibnamefont{McDermott}}, \bibinfo{journal}{Phys. Rev. Lett.}
  \textbf{\bibinfo{volume}{123}}, \bibinfo{pages}{191102}
  (\bibinfo{year}{2019}), \eprint{1905.02727}.

\bibitem[{\citenamefont{Huang et~al.}(2018)\citenamefont{Huang, Ohlsson, and
  Zhou}}]{Huang:2017egl}
\bibinfo{author}{\bibfnamefont{G.-y.} \bibnamefont{Huang}},
  \bibinfo{author}{\bibfnamefont{T.}~\bibnamefont{Ohlsson}}, \bibnamefont{and}
  \bibinfo{author}{\bibfnamefont{S.}~\bibnamefont{Zhou}},
  \bibinfo{journal}{Phys. Rev. D} \textbf{\bibinfo{volume}{97}},
  \bibinfo{pages}{075009} (\bibinfo{year}{2018}), \eprint{1712.04792}.

\bibitem[{\citenamefont{Venzor et~al.}(2021)\citenamefont{Venzor,
  P\'erez-Lorenzana, and De-Santiago}}]{Venzor:2020ova}
\bibinfo{author}{\bibfnamefont{J.}~\bibnamefont{Venzor}},
  \bibinfo{author}{\bibfnamefont{A.}~\bibnamefont{P\'erez-Lorenzana}},
  \bibnamefont{and}
  \bibinfo{author}{\bibfnamefont{J.}~\bibnamefont{De-Santiago}},
  \bibinfo{journal}{Phys. Rev. D} \textbf{\bibinfo{volume}{103}},
  \bibinfo{pages}{043534} (\bibinfo{year}{2021}), \eprint{2009.08104}.



\bibitem[{\citenamefont{Lorenz et~al.}(2021)\citenamefont{Lorenz, Funcke,
  L\"offler, and Calabrese}}]{Lorenz:2021alz}
\bibinfo{author}{\bibfnamefont{C.~S.} \bibnamefont{Lorenz}},
  \bibinfo{author}{\bibfnamefont{L.}~\bibnamefont{Funcke}},
  \bibinfo{author}{\bibfnamefont{M.}~\bibnamefont{L\"offler}},
  \bibnamefont{and}
  \bibinfo{author}{\bibfnamefont{E.}~\bibnamefont{Calabrese}},
  \bibinfo{journal}{Phys. Rev. D} \textbf{\bibinfo{volume}{104}},
  \bibinfo{pages}{123518} (\bibinfo{year}{2021}), \eprint{2102.13618}.

\bibitem[{\citenamefont{Boehm and Fayet}(2004)}]{Boehm:2003hm}
\bibinfo{author}{\bibfnamefont{C.}~\bibnamefont{Boehm}} \bibnamefont{and}
  \bibinfo{author}{\bibfnamefont{P.}~\bibnamefont{Fayet}},
  \bibinfo{journal}{Nucl. Phys. B} \textbf{\bibinfo{volume}{683}},
  \bibinfo{pages}{219} (\bibinfo{year}{2004}), \eprint{hep-ph/0305261}.



\end{thebibliography}

\end{document}